\documentclass[onecolumn, 11pt,letterpaper]{article}

\usepackage{epsfig}
\usepackage{subfigure}
\usepackage{amsmath}
\usepackage{float}
\usepackage{amssymb}
\usepackage{graphicx}
\usepackage{algorithm}
\usepackage{algorithmic}
\usepackage[left=1in,top=1in,right=1in,bottom = 1in, nohead]{geometry}

\newcommand{\eat}[1]{}

\newtheorem{theorem}{Theorem}
\newtheorem{lemma}{Lemma}

\newtheorem{corollary}{Corollary}
\newtheorem{definition}{Definition}
\newtheorem{example}{Example}

\def\QEDopen{{\setlength{\fboxsep}{0pt}\setlength{\fboxrule}{0.2pt}\fbox{\rule[0pt]{0pt}{1.3ex}\rule[0pt]{1.3ex}{0pt}}}}
\def\QED{\QEDopen}
\def\proof{\noindent{\bf Proof}: }
\def\endproof{\hspace*{\fill}~\QED\par\endtrivlist\unskip}

\newcommand{\Vector}[1]{\mbox{\boldmath$#1$}}

\newcommand{\vx}{\Vector{x}}
\newcommand{\vsigma}{\Vector{\sigma}}
\newcommand{\vu}{\Vector{u}}
\newcommand{\vA}{\Vector{A}}
\newcommand{\vv}{\Vector{v}}
\newcommand{\vy}{\Vector{y}}

\newcommand{\vw}{\Vector{w}}
\newcommand{\vM}{\Vector{M}}
\newcommand{\vm}{\Vector{m}}

\newcommand{\ve}{\Vector{e}}

\newcommand{\vD}{\Vector{D}}

\newcommand{\vpsi}{\Vector{\psi}}
\newcommand{\vPsi}{\Vector{\Psi}}

\newcommand{\vPhi}{\Vector{\Phi}}

\newcommand\qip[2]{\langle #1 , #2 \rangle}

\long\def\comment#1{}
\date{}
\begin{document}
\title{Compressive Mechanism: Utilizing Sparse Representation in Differential Privacy}


\author{
Yang D. Li$^1$ \and Zhenjie Zhang$^1$ \and  Marianne Winslett$^{1,2}$ \and Yin Yang$^1$  \\
\and $^1$Advanced Digital Sciences Center\\ Illinois at Singapore Pte.\\
\{daniel.li, zhenjie, yin.yang\}@adsc.com.sg\\
\and $^2$Department of Computer Science\\ University of Illinois at Urbana-Champaign\\
winslett@illinois.edu\\
}

\date{}

\maketitle 
\begin{abstract}
Differential privacy provides the first theoretical foundation
with provable privacy guarantee against adversaries with arbitrary
prior knowledge. The main idea to achieve differential privacy is
to inject random noise into statistical query results. Besides
correctness, the most important goal in the design of a
differentially private mechanism is to reduce the effect of random
noise, ensuring that the noisy results can still be useful.

This paper proposes the \emph{compressive mechanism}, a novel
solution on the basis of state-of-the-art compression technique,
called \emph{compressive sensing}. Compressive sensing is a decent
theoretical tool for compact synopsis construction, using random
projections. In this paper, we show that the amount of
noise is significantly reduced from $O(\sqrt{n})$ to $O(\log(n))$,
when the noise insertion procedure is carried on the synopsis
samples instead of the original database. As an extension, we also
apply the proposed compressive mechanism to solve the problem of
continual release of statistical results. Extensive experiments
using real datasets justify our accuracy claims.
\end{abstract}

\section{Introduction}\label{sec:intro}

\small {\it No one shall be subjected to arbitrary interference
with his privacy, family, home or correspondence, nor to attacks
upon his honor and reputation. }
\begin{flushright} Universal Declaration of Human Rights
\end{flushright}
\normalsize

Rapid advances in information technology and computational capacity are raising concerns regarding the privacy of sensitive personal information. Previous work has shown that even after removing all personal identity attributes such as names and addresses, adversaries may still be able to identify specific individuals by combining their \emph{prior knowledge} with the published attributes in the database, e.g., age, gender and race. For example, in 2007, a team of researchers successfully reidentified two customers in the ``anonymized'' Netflix data set \cite{netflix}, based on their transaction histories with Netflix and movie comments on IMDB \cite{NS08}. Moreover, recent studies show that group statistics are also vulnerable to privacy attacks due to inference techniques combining anonymized data and existing and/or public information. In Genome Wide Association Studies (GWAS), for instance, the DNA samples from participating patients are mixed to prevent disclosure of their identities, and only statistics regarding the prevalence of particular single-nucleotide polymorphisms (SNPs) are published. However, an adversary can still verify the presence of a specific person in the mixture with high confidence, given a DNA sample from that person and a reference DNA mixture \cite{HSR+08,WLW+09}. Such privacy
problems have become a major obstacle to biomedical research, as they make access to useful input data increasingly difficult to obtain.

To tackle these problems, numerous privacy protection frameworks have been proposed. Among these, \emph{differential privacy} outshines others in providing strong robustness guarantee against attacks from adversaries with arbitrary prior knowledge. Simply put, a randomized mechanism (query answering method) for answering statistical queries satisfies differential privacy if and only if the query results will be almost identical after modifying or deleting one of the records in the database \cite{DBLP:conf/tcc/DworkMNS06}. This requirement ensures that sensitive information in a record almost cannot be inferred, even if the adversary knows the values of all the other records in the database.

To achieve differential privacy, one basic mechanism is to add random noise to the statistical query results \cite{DBLP:conf/tcc/DworkMNS06}. In particular, given the unperturbed query result and a parameter $\epsilon$ called the \emph{privacy budget}, the mechanism randomly selects a number following a \emph{Laplacian distribution}, with mean at the original query result and scale proportional to the \emph{sensitivity} (discussed later) of the queries times $\frac{1}{\epsilon}$. Theoretical analysis shows that the resulting randomness in the query result fulfills the requirements of differential privacy with respect to privacy budget $\epsilon$. While this basic mechanism handles a single query quite well, it is not effective for processing a large number of queries, as it requires a privacy budget linear in the number of queries to maintain a given level of result accuracy. Thus queries may quickly exhaust the total privacy budget assigned to a database, forcing the database to be taken offline to avoid unacceptable privacy breaches. In such situations, a better alternative is to build a compact, privacy-preserving synopsis from the data with a fixed amount of privacy budget, such that it is capable of answering all possible queries.

There have been several previous efforts to design and construct such synopses, e.g., wavelets \cite{DBLP:conf/icde/XiaoWG10}, trees \cite{DBLP:journals/pvldb/HayRMS10} and linear summation basis \cite{DBLP:conf/pods/LiHRMM10,DBLP:conf/stoc/DworkNPR10,DBLP:conf/icalp/ChanSS10}. These synopsis structures are \emph{lossless}, meaning that they preserve all information in the original database. In other words, the original statistics can be completely and accurately recovered by running the corresponding decoding algorithms on the synopsis structures. Consequently, the size of these synopses, as well as the privacy budget they require, grow linearly with the size of the dataset.

\begin{figure}[t]
\centering
\begin{tabular}[t]{c}
   \subfigure{
      \psfig{figure=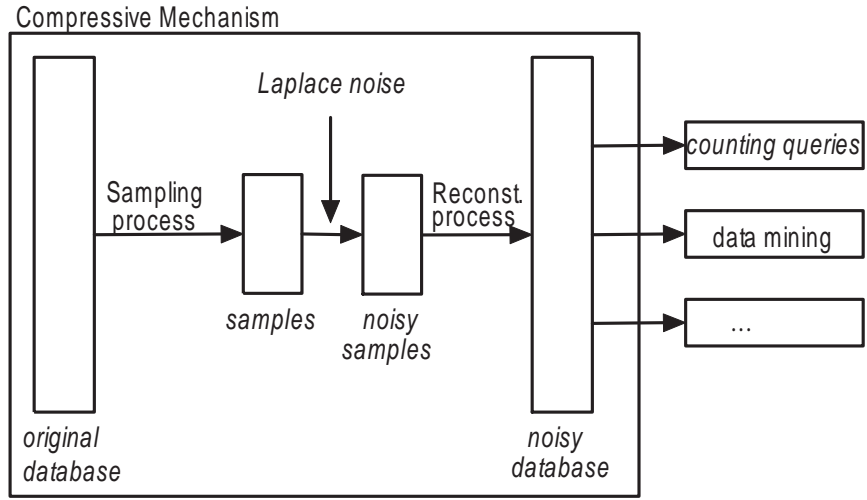,width=0.7\linewidth}
      \label{diagram}
   }
\end{tabular}
\vspace{-5pt} \caption{\label{Figure 1} Compressive
mechanism framework.} \vspace{-15pt} \label{fig: diagram}
\end{figure}

This paper explores a new direction: probabilistic synopses
based on \emph{compressive sensing}
\cite{DBLP:journals/tit/Donoho06,CW2008,DBLP:journals/tit/CandesT05}.
Figure \ref{fig: diagram} illustrates
this new compressive sensing mechanism. Using a \emph{sparse representation} of the
original data, we use compressive sensing to encode a very small synopsis, compared to the original database size.  We then add Laplacian noise to the synopsis, making it differentially private; decode the synopsis, creating a noisy version of the original data; then answer an unlimited number of queries over the decoded data, without adding additional noise. The compressive sensing mechanism allows us to use less noise than previous synopsis
proposals under certain conditions, and provides much more
accurate statistical query results after decoding. Unlike previous
methods that focus on specific classes of queries, the compressive sensing
mechanism is \emph{universal}, supporting all possible queries
on the decoded noisy data. Thus the compressive sensing
mechanism can be seamlessly incorporated into any applications
with privacy concerns, from GWAS analysis to
user transaction history mining. We show that the compressive mechanism improves the accuracy of the result
statistics by up to an order of magnitude, in both theoretical
analysis as well as empirical studies.

The organization of the paper is as follows. Section \ref{sec:related} summarizes existing studies of differential privacy and compressive sensing. Section \ref{sec:prelim} gives an overview of differential privacy and compressive sensing. Section \ref{sec:definition} introduces \emph{universal mechanisms} and the basic definitions of our target problems. Section \ref{sec:mechanism} presents the compressive mechanism and analyzes its error. Section \ref{sec:continual} generalizes the compressive mechanism to a streaming environment, for continual release of statistics. Section \ref{sec:exp} demonstrates the usefulness of these new methods through empirical studies with real data sets. Finally, Section \ref{sec:conclu} concludes the paper and outlines interesting research directions for future work.

\section{Related Work}\label{sec:related}

This section gives an overview of work on differential privacy and compressive sensing.

\subsection{Differential Privacy}

The notion of $\epsilon$-differential privacy was introduced in \cite{DBLP:conf/tcc/DworkMNS06, DBLP:conf/icalp/Dwork06}. Alternative definitions soon appeared, such as $(\epsilon, \delta)$-differential privacy \cite{DBLP:conf/eurocrypt/DworkKMMN06, DBLP:conf/stoc/RothR10}, pan-privacy \cite{DBLP:conf/innovations/DworkNPRY10, DBLP:conf/stoc/DworkNPR10} and zero-knowledge privacy \cite{DBLP:conf/tcc/GehrkeLP11}. $(\epsilon, \delta)$-differential privacy relaxes $\epsilon$-differential privacy, while pan-privacy and zero-knowledge privacy are stronger than $\epsilon$-differential privacy and apply to special circumstances. This work uses the more popular $\epsilon$-differential privacy definition. \cite{DBLP:journals/cacm/Dwork11} provides a comprehensive survey of the development of differential privacy.

Considerable efforts have been devoted to the design of $\epsilon$-differentially private mechanisms, but the majority of them only deal with linear counting queries, e.g., \cite{DBLP:conf/icalp/ChanSS10, DBLP:conf/icde/XiaoWG10, DBLP:conf/stoc/HardtT10, DBLP:conf/pods/LiHRMM10, DBLP:conf/stoc/DworkNPR10, DBLP:conf/stoc/GhoshRS09, DBLP:journals/pvldb/HayRMS10}. In contrast, our compressive sensing mechanism can handle \emph{arbitrary} statistical queries. Particularly, \cite{DBLP:conf/icde/XiaoWG10} uses a Haar wavelet transform and \cite{DBLP:journals/pvldb/HayRMS10} designs a tree structure for answering range counting queries. However, neither the Haar wavelet transform nor tree structure can be generalized to answer other queries in a private and accurate way. \cite{DBLP:conf/icde/XiaoWG10} and \cite{DBLP:journals/pvldb/HayRMS10} are merged into a unified framework of \cite{DBLP:conf/pods/LiHRMM10} for answering linear counting queries.  A recent work \cite{De2011} separates counting queries from arbitrary low-sensitivity queries and essentially shows that counting is much easier than arbitrary low-sensitivity queries. \cite{DBLP:conf/sigmod/RastogiN10, DBLP:conf/pods/BarakCDKMT07} use (deterministic) Fourier transforms, for problems that are very different from ours; while we take advantage of a probabilistic transform in compressive sensing to reduce the dimension of data from $O(n)$ to $O(\log n)$.

\subsection{Compressive Sensing}

Compressive sensing was introduced by \cite{DBLP:journals/tit/CandesRT06, DBLP:journals/tit/CandesT06, DBLP:journals/tit/Donoho06}, and later shown to have extensive applications in imaging \cite{Sankaranarayanan:2010:CAD:1886063.1886075, LDP2007, BS2007}, signal processing \cite{TWDBB2006}, computational biology \cite{SSMB2007}, geophysical data analysis \cite{LH2007}, communications \cite{TH2008} and so on. To the best of our knowledge, we are the first to apply compressive sensing to sensitive data analysis.

\cite{Zhou:2009:DPC:1700967.1701079, DBLP:journals/tit/ZhouLW09, DBLP:journals/tkde/LiuKR06} apply random projections to differential privacy. They show that the compressed data can be used for certain statistical tasks, and do not consider the reconstruction process of compressive sensing. \cite{DBLP:journals/corr/abs-1103-0825} reconsiders the contingency table release problem \cite{DBLP:conf/pods/BarakCDKMT07} for sparse data (without any transformation), without using compressive sensing. \cite{DBLP:conf/stoc/DworkMT07} is the most relevant paper in the sense that it works in the opposite direction of ours: applying privacy to compressive sensing.

\section{Preliminaries}\label{sec:prelim}

\subsection{Notation}

We use $\mathbb{R}$ ($\mathbb{R}^+$) to denote the set of real numbers (positive real numbers) and $[1,n]$ to represent the integer set $\{1, 2, \ldots, n\}$. For two vectors $\vv, \vw \in \mathbb{R}^n$, $\qip{\vv}{\vw}$ means their inner product. The $p$-norm ($p$ is a positive integer) of a vector $\vv \in \mathbb{R}^n$ is defined to be $(\sum_{i = 1}^n |\vv[i]|^p)^{\frac{1}{p}}$ and is denoted by $||\vv||_p$. For example, the 2-norm of the difference of two points gives their Euclidian distance. We write $\log$ for $\log_2$. The number $e$ is the base of natural logarithms. We let $Lap(\lambda)$ denote the one-dimensional Laplacian distribution centered at $0$ with scale $\lambda$ and the corresponding density function $g(x) = \frac{1}{2\lambda}e^{-\frac{|x|}{\lambda}}$. The composition of functions $g$ and $h$ is denoted $g \circ h$, meaning that we first apply $g$ to input and then $h$ to the output of $g$.

We use other notation common in theoretical computer science. When describing asymptotic complexity, we use $\tilde{O}$ (pronounced soft-$O$) as a variant of $O$ (pronounced big-$O$) that ignores logarithmic factors. For instance, if the complexity is $O(n/\log n)$, we simply write $\tilde{O}(n)$. Also, $\exp(n)$ means $e^{Cn}$ for some constant $C$. \emph{With high probability} means with probability at least $0.99$.

\subsection{Compressive Sensing}
This section gives a brief overview of the theory of compressive sensing, and we refer the reader to an excellent survey \cite{CW2008} for more information. As shown in Figure \ref{fig: diagram}, compressive sensing consists of a probabilistic compression procedure, also called the \emph{sampling} process, followed by a \emph{reconstruction} process that decodes the compressed data. The sampling process reduces the data size from $O(n)$ to $O(\log n)$.
The rather complex decoding process exactly or approximately reconstructs the original data from the compressed samples.
Readers who are not interested in the mathematics underlying the compressive sensing technique should skip to the next section.

In what follows, all vectors are over $\mathbb{R}^n$ unless otherwise noted. Consider a vector $\vD$ that we wish to represent using an \emph{orthonormal} basis (such as a standard basis, or a wavelet basis) $\vPsi = [\vpsi_1,\ldots , \vpsi_n]$.  Let $\vx = (\vx[1], \ldots, \vx[n])$ be the coefficient sequence of $\vD$ under the new basis $\vPsi$. Then we have:
\begin{align*}
\vD[j] = \sum_{i=1}^n\vx[i]\vpsi_i[j].
\end{align*}
If we treat $\vPsi$ as an $n \times n$ matrix with $\vpsi_1, \ldots, \vpsi_n$ as the columns, $\vD$ can be written as $\vPsi \vx$, and we say that $\vx$ represents $\vD$ under the new basis. We call $\vx$ \emph{$S$-sparse} if it has at most $S$ nonzero entries. Let $\vx_S$ be obtained from $\vx$ by replacing its $n - S$ coefficients with smallest absolute value by 0. Then $\vx_S$ is $S$-sparse.

Often data are compressible. Given a constant $0<p<1$, we say vector $\vx$ is \emph{($p$-)compressible} with magnitude $R$ if its components taken in sorted order obey $|\vx_{(i)}| \le R\cdot i^{-1/p}, \forall i \in [1,n]$. A compressible vector $x$ can be well approximated by an $S$-sparse vector in the sense that $||\vx - \vx_S||_1 \le C_p\cdot R\cdot S^{1-1/p}$ for some constant $C_p$. A vector $\vD$ has an \emph{($S$-)sparse representation} if there is an orthonormal basis $\vPsi$ (called a \emph{sparse basis}) where $\vD$'s representation $\vx$ is sparse or compressible, i.e., $\vx$ is $S$-sparse or is compressible to an $S$-sparse vector $\vx_S$.

The input to compressive sensing is a vector $\vD$ with a sparse representation. Then there is a sampling process which can be characterized as a linear mapping. We use a matrix $\vPhi \in \mathbb{R}^{k\times n}$ to describe the sampling operator and the result is a vector $\vy = \vPhi\vD \in \mathbb{R}^k$. Candes and Tao \cite{DBLP:journals/tit/CandesT05} define the $r$-th \emph{restricted isometry constant} $\delta_r$ of $\vPhi$ to be the smallest number such that
\begin{align*}
(1 - \delta_r)||\vx||_2^2 \le ||\vPhi\vx||_2^2 \le (1 + \delta_r)||\vx||_2^2,
\end{align*}
for all $r$-sparse vectors $\vx \in \mathbb{R}^n$. The condition that $\delta_{2S} \ll 1$ implies that all pairwise distances between $S$-sparse signals must be well-preserved in the measurement space, i.e.,
\begin{align*}
(1 - \delta_{2S})||\vx_1 - \vx_2||_2^2 \le ||\vPhi\vx_1 - \vPhi\vx_2||_2^2 \le (1 + \delta_{2S})||\vx_1 - \vx_2||_2^2,
\end{align*}
for all $S$-sparse vectors $\vx_1, \vx_2$. We loosely say that a matrix $\vPhi$ satisfies the \emph{Restricted Isometry Property} (RIP) if it has $\delta_{2S} \ll 1$. Such $\vPhi$ always exists. For example, with probability $1 - \exp(- k)$, a random matrix $\vPhi$ formed by sampling independent and identically distributed (i.i.d.) entries from a symmetric Bernoulli distribution, more precisely
\begin{align*}
Prob(\vPhi(i,j) = \pm \frac{1}{\sqrt{k}}) = \frac{1}{2},
\end{align*} satisfies RIP, provided that
\begin{align*}
k \ge C\cdot S\log(n/S),
\end{align*}
where $C$ is some constant \cite{BDDW2008}. In words, every entry of $\vPhi$ equals $\frac{1}{\sqrt{k}}$ with probability $1/2$ and equals $- \frac{1}{\sqrt{k}}$ with probability $1/2$. In the following discussion, we will assume that $\vPhi \in \mathbb{R}^{k\times n}$ is in such a form. It is not hard to verify that $\vA = \vPhi \vPsi \in \mathbb{R}^{k\times n}$ also satisfies RIP with overwhelming probability $1 - \exp(- k)$, where $k = \Theta(S\log(n/S))$, for any fixed orthonormal basis $\vPsi \in \mathbb{R}^{n\times n}$ \cite{BDDW2008}.

Up to now, we have clarified the input and the sampling process, obtaining a sample vector $\vy = \vPhi\vD \in \mathbb{R}^k$ with $k = \Theta(S\log(n/S))$. The next and final step is to reconstruct the vector $\vD$ from $\vy$ through the sparse representation of $\vD$. The samples may be contaminated with an unknown noise $\ve \in \mathbb{R}^k$, and the sample vector becomes
\begin{align*}
\vy^* = \vy + \ve = \vPhi\vD + \ve = \vA\vx + \ve,
\end{align*}
where $\vA = \vPhi \vPsi$ is known from the sampling process. Candes, Romberg and Tao \cite{CRT2006} prove the remarkable and surprising result that by solving a combinatorial optimization problem, the recovered answer $\vx^* \in \mathbb{R}^n$ can be close enough to $\vx$ even in the presence of unknown perturbations. Needell and Tropp \cite{DBLP:journals/cacm/NeedellT10} prove essentially the same error bound using a greedy algorithm. The details of the two algorithms can be found in the appendix and the result is summarized as follows.
\begin{lemma}[\cite{CRT2006, DBLP:journals/cacm/NeedellT10}]\label{CRTresult}
Suppose $\vA$ satisfies RIP and $||\ve||_2 \le \theta$. Then $||\vx - \vx^*||_2 \le \frac{C_2||\vx - \vx_S||_1}{\sqrt{S}} + C_3\theta$, for some constants $C_2$ and $C_3$.
\end{lemma}

For a compressible vector $\vx \in \mathbb{R}^n$, we mentioned that $||\vx - \vx_S||_1 \le C_p\cdot R\cdot S^{1-1/p}$ for some $p \in (0, 1)$. Also note that $\vD$ can be recovered by setting $\vD^* = \vPsi\vx^*$ and that $||\vD - \vD^*||_2 = ||\vx - \vx^*||_2$. We can derive the following corollary for our purpose.
\begin{corollary}\label{ErrorBound}
Suppose that $\vA$ satisfies RIP and $||\ve||_2 \le \theta$.  Then $||\vD - \vD^*||_2 = O(S^{1/2-1/p} + \theta)$ for some constant $S \ll n$ and $p \in (0,1)$.
\end{corollary}

If the input vector $\vD \in \mathbb{R}^n$ has no sparse representation, then the $\vD^* \in \mathbb{R}^n$ obtained by compressive sensing could be very far away from $\vD$, and that is why we require the input to have a sparse representation. Formally,
\begin{corollary}\label{generalbound}
Suppose $\vA$ satisfies RIP and $||\ve||_2 \le \theta$.  If $\vD$ has no sparse representation, then $||\vD - \vD^*||_2 = O(n/\sqrt{S} + \theta)$.
\end{corollary}

Finally, we remark that the time complexity and the space complexity of compressed sensing are both $\tilde{O}(n)$.

\subsection{Differential Privacy}
In this paper, we represent a database as a vector $\vD \in \mathbb{R}^n$. This abstraction encompasses many previous abstractions of data, such as a data distribution \cite{DBLP:conf/stoc/RothR10}, histogram \cite{DBLP:conf/stoc/HardtT10}, contingency table \cite{DBLP:conf/pods/BarakCDKMT07}, private bits \cite{DBLP:conf/pods/DinurN03}, the database itself, and recommendation systems \cite{DBLP:conf/kdd/McSherryM09}. Two databases $\vD_1, \vD_2 \in \mathbb{R}^n$ are said to be \emph{neighboring} iff $||\vD_1 - \vD_2||_1 \le 1$. The notion of differential privacy is defined as follows.
\begin{definition}[\cite{DBLP:conf/tcc/DworkMNS06, DBLP:conf/icalp/Dwork06}]\label{DP}
A randomized mechanism $\mathcal{K}$ provides $\epsilon$-differential privacy if for all neighboring databases $\vD_1, \vD_2 \in \mathbb{R}^n$ and all $Sub \subseteq Range(\mathcal{K})$,
\begin{align*}
Prob(\mathcal{K}(\vD_1) \in Sub) \le e^\epsilon \times Prob(\mathcal{K}(\vD_2) \in Sub),
\end{align*}
where the probability space in each case is over the coin flips of $\mathcal{K}$.
\end{definition}

A popular mechanism for achieving $\epsilon$-differential privacy is the \emph{Laplacian mechanism} \cite{DBLP:conf/tcc/DworkMNS06}, which can be used when the output of the mechanism is numeric. McSherry and Talwar developed a technique called the \emph{exponential mechanism} \cite{DBLP:conf/focs/McSherryT07} for problems where the output is non-numeric.

\emph{Laplacian mechanism: } The \emph{sensitivity} of a query $Q: \mathbb{R}^n \rightarrow \mathbb{R}^d$ is defined to be
\begin{align*}
\Delta_Q = \max_{\vD_1, \vD_2} ||g(\vD_1) - g(\vD_2)||_1,
\end{align*}
for all neighboring $\vD_1, \vD_2 \in \mathbb{R}^n$. \cite{DBLP:conf/tcc/DworkMNS06} shows the following result:
\begin{lemma}[\cite{DBLP:conf/tcc/DworkMNS06}]\label{LaplaceMechanism}
For $Q: \mathbb{R}^n \rightarrow \mathbb{R}^d$, the mechanism $\mathcal{K}_Q$ that adds independently generated noise with distribution $Lap(\Delta_Q/\epsilon)$ to each of the $d$ output values provides $\epsilon$-differential privacy.
\end{lemma}

\emph{Exponential mechanism: }This mechanism is for the case where the query answer $y$ is not numerical.  We rely on a pre-defined utility function $u(\vD, y)$ (with a numeric output) to measure the quality of $y$, compared to the exact answer. The exponential mechanism outputs $y$ with probability proportional to $e^{-\epsilon u(\vD, y)/(2\Delta u)}$, where $\Delta u$ is the sensitivity of the utility function $u(\vD, y)$. The exponential mechanism provides $\epsilon$-differential privacy \cite{DBLP:conf/focs/McSherryT07}. The distance of an answer from the best answer, which has the smallest $u$, exhibits an exponential tail and with probability almost $1$, the exponential mechanism outputs an object with an approximately optimal value.

\emph{Continual mechanism and pan-privacy: }For the case where the database is updated over time, the theory community has investigated what they call differential privacy under continual observation \cite{DBLP:conf/stoc/DworkNPR10}. In their setting, the input is no longer a static vector, but instead a stream of $0$'s and $1$'s, denoted by $\vsigma \in \{0, 1\}^T$, where $T$ is an upper limit of time. The \emph{continual mechanism} \cite{DBLP:conf/stoc/DworkNPR10} receives an input $\vsigma[t] \in \{0,1\}$ at each time $t \in [1, T]$, and outputs an approximation to the number of $1$'s seen in the length $t$ prefix of the stream. Two streams' \emph{prefixes} $\vsigma_t, \vsigma_t' \in \{0, 1\}^t$, $t \in [1, T]$, are \emph{neighboring} iff $||\vsigma_t - \vsigma'_t||_1 \le 1$. The definition of \emph{$\epsilon$-differential pan-privacy} under continual observation \cite{DBLP:conf/innovations/DworkNPRY10, DBLP:conf/stoc/DworkNPR10} is stronger than $\epsilon$-differential privacy and is stated as follows.
\begin{definition}[\cite{DBLP:conf/innovations/DworkNPRY10, DBLP:conf/stoc/DworkNPR10}]
Let $I_{\mathcal{K}}$ denote the set of internal states of the randomized mechanism $\mathcal{K}$. $\mathcal{K}$ provides $\epsilon$-differential pan-privacy (against a single intrusion) if for all neighboring stream prefixes $\vsigma_t, \vsigma_t' \in \{0, 1\}^t$, $t \in [1,T]$ and for all sets $I' \subseteq I_{\mathcal{K}}$ and $Sub \subseteq Range(\mathcal{K})$,
\begin{align*}
Prob(\mathcal{K}(\vsigma_t) \in (I', Sub)) \le e^\epsilon \times Prob(\mathcal{K}(\vsigma'_t) \in (I', Sub)),
\end{align*}
where the probability space in each case is over the coin flips of $\mathcal{K}$.
\end{definition}

The details of the continual mechanism are included in the appendix. The continual mechanism can be easily generalized to the case in which input $\vD \in \mathbb{R}^T$. If we denote the approximate sum of the first $t$ entries by $\Sigma_t^*$ and the true sum of the first $t$ input items as $\Sigma_t$, the result of \cite{DBLP:conf/stoc/DworkNPR10} can be formalized (in a slightly differently way) as follows.
\begin{lemma}[\cite{DBLP:conf/stoc/DworkNPR10}]\label{continual}
The continual mechanism provides $\epsilon$-\\ differential pan-privacy. At each time $t \in [1, T]$, with probability at least $1 - \beta$, $|\Sigma_t^* - \Sigma_t| = O(\log(1/\beta)\log^{1.5}(T)/\epsilon)$.
\end{lemma}

\section{Problem Definition}\label{sec:definition}

This section formally defines the problems investigated in this paper. Specifically, we focus on \emph{$\epsilon$-differentially private randomized mechanisms} with \emph{numeric} outputs, which can be formalized as $\mathcal{K}: \mathbb{R}^n \rightarrow \mathbb{R}^d$. Such randomized mechanisms can be used to answer (numeric) statistical queries (in the form of $Q: \mathbb{R}^n \rightarrow \mathbb{R}^d$) about a database. In general, a randomized mechanism for publishing query results about a database must resolve a trade-off between \emph{utility} and \emph{privacy}. Utility means that the outputs should not be too far away from the true answers of the query, to ensure that the perturbed answers can still be helpful to users.  Privacy requires that the outputs not be too near to the true answers, since some amount of random perturbation is essential for the mechanism to be $\epsilon$-differentially private. While many previous studies mainly deal with \emph{linear counting} queries, in practice the queries are likely to be much more diverse and complicated, such as a multiphase analysis task. A \emph{general} randomized mechanism with respect to a certain kind of database should be able to answer \emph{all possible} queries with guarantees about utility and privacy. In other words, given a database $\vD$ of a certain kind $X$, any query $Q$ over $\vD$ has to be answered with reasonable utility and privacy guarantees. For this purpose, this section defines the notion of a \emph{universal mechanism}. We begin with the notion of an \emph{identity query}, which just returns the entire database unchanged. Intuitively, the purpose of a universal mechanism is to accurately answer the identity query, subject to differential privacy.

\begin{definition}
A universal mechanism with respect to a class $X$ of databases is a randomized mechanism $\mathcal{U}_X: X \rightarrow \mathbb{R}^n$ for answering the identity query, satisfying the following conditions:
\begin{itemize}
\item $\epsilon$-differential privacy;
\item with high probability, $||\vD - \vD^*||_2 = O(\log(n)/\epsilon)$, for any input $\vD \in X$ and its corresponding output $\vD^* \in \mathbb{R}^n$.
\end{itemize}
\end{definition}

A universal mechanism is a good base for answering all kinds of statistical queries. To answer any query $Q$, we first apply $\mathcal{U}_X$ to a database $\vD \in X$, obtaining $\vD^*$.  Then we deterministically compute an answer to $Q$ over $\vD^*$. Since $\vD^*$ must be very close to $\vD$, utility can be guaranteed for any query.

Another important property of a universal mechanism is that it allows us to answer an unbounded number of queries without \emph{any concern} for privacy budget issues. Previous $\epsilon$-differential privacy mechanisms can only answer a finite number of statistical queries, due to the budget limit. For example, if a mechanism answers two queries each in an $\epsilon$-differentially private way, then the mechanism may have provided only $2\epsilon$-differential privacy overall. The universal mechanism does not have this problem, as any system based on a universal mechanism \emph{always} satisfies $\epsilon$-differential privacy, no matter how many queries are asked.

Another advantage of universal mechanisms is that $\vD^*$ can be published in its entirety, thus supporting both interactive and non-interactive querying. For example, biologists who perform GWAS can simply publish allele frequencies, rather than answering queries about the frequencies. So a universal mechanism itself can be very useful, even without consideration of subsequent queries.

We use $O(\log(n)/\epsilon)$ as the upper bound of error in the
definition of universal mechanism. One reason for this choice is
that we conjecture that $\Omega(\log(n)/\epsilon)$ is the lower
bound of error to satisfy $\epsilon$-differential privacy.
Previous lower bounds \cite{DBLP:conf/stoc/HardtT10,
DBLP:conf/pods/DinurN03, De2011} cannot be applied to our case,
and therefore we leave this conjecture as an open problem,
elaborated further at the end of this article.

To summarize, a universal mechanism is resilient to \emph{any form} and \emph{any number} of statistical attacks from any number and kind of \emph{malicious attackers}, which makes it \emph{universally robust}. Meanwhile, the statistical query results remain relatively accurate, as $\vD^*$ must be very close to $\vD$.

The Laplacian mechanism is not a universal mechanism, as it introduces too much error. Suppose we
use it to answer the identity query for $\vD$, producing $\vD^*$. With high probability $||\vD - \vD^*||_2 =
\Theta(\sqrt{n}/\epsilon)$ (by using a Chernoff-like argument).
Nor is any
other known $\epsilon$-differentially private randomized mechanism
universal. The challenge, then is to devise a universal mechanism.  In later sections we introduce the  \emph{compressive
mechanism} and show that it is a universal mechanism with respect to databases with a
sparse representation. Table \ref{tab:node} compares
the error bounds of the compressive mechanism and other
contenders. In the remainder of this section, we present two example use cases
for universal mechanisms.
\begin{table}[t]
\centering \caption{Error bounds (in terms of $||\vD - \vD^*||_2$)
for the identity query.}
\begin{tabular}{c c}
\hline \hline
Mechanism&   Bound \\
\hline
 \vspace{1mm} Compressive Mechanism & $O(\log n/\epsilon)$\\
 \vspace{1mm} Laplacian Mechanism & $O(\sqrt{n}/\epsilon)$  \\
 \vspace{1mm} \cite{DBLP:conf/icde/XiaoWG10, DBLP:journals/pvldb/HayRMS10}& $\tilde{O}(\sqrt{n}/\epsilon)$\\
\hline
\end{tabular}\label{tab:node}
\end{table}

\begin{example}
Biomedical researchers use human genome sequence data to determine
the correlations between particular
diseases and combinations of SNPs. They perform
statistical tests on the data after computing preliminary
estimates of the importance of the SNPs \cite{WLW+09}.
In the US, frequency information for the SNPs in NIH-funded studies is no longer publicly available, due to privacy concerns.  A universal mechanism could allow publication of this information, which would be used by many other researchers.
\end{example}

\begin{example}
Internet service providers (ISPs) share statistical data from
their network traces, to help detect anomaly events of
large scale \cite{NJC+10}. The identification of anomalies involves
matrix computations on the statistics, which are not generally
supported by existing mechanisms for differential privacy.
\end{example}

A universal mechanism can also be helpful for databases that get updated. The continual mechanism of
\cite{DBLP:conf/stoc/DworkNPR10} can only answer linear counting
queries (more precisely, only range counting queries starting from
time $1$). We aim to establish a pan-private mechanism such that
at each time $t$, we can answer any
query over the database states ($\vD[1], \ldots,
\vD[t]$) we have seen so far. We will realize this ambition by
designing a mechanism for answering the identity query at each
time $t$. The definition of the identity query in a dynamic setting
is a straightforward generalization of identity queries in a
static setting, and we omit it here.

\begin{figure}[t]
\centering
\begin{tabular}[t]{c}
   \subfigure{
      \psfig{figure=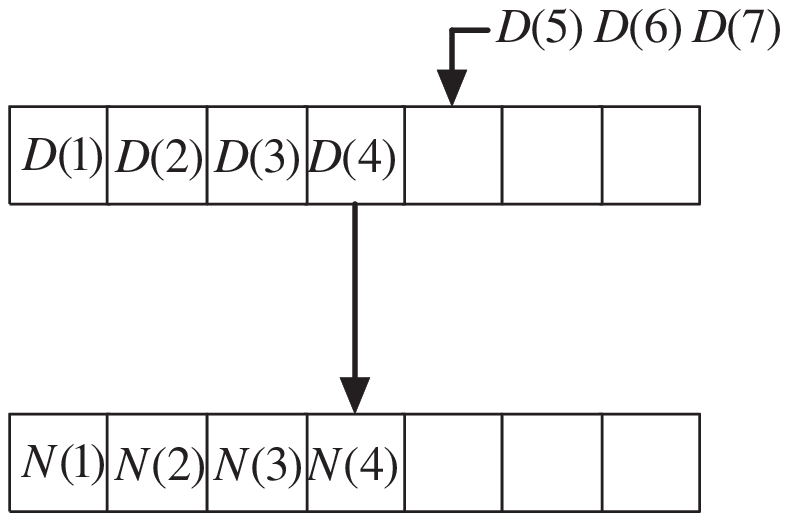,width=0.5\linewidth}
      \label{diagram}
   }
\end{tabular}
\vspace{-5pt} \caption{A universal mechanism for databases with updates
(continual observation).} \vspace{-15pt} \label{fig: continual}
\end{figure}

Figure \ref{fig: continual} illustrates how a universal mechanism
works under continual observation. When a new tuple $D(4)$ comes
into the database, an updating procedure is triggered to
generate a corresponding noisy tuple $N(4)$ under a universal mechanism.
The noisy values calculated so far are published
continuously to the public, guaranteeing that any snapshot of the
publication follows the requirements of a universal mechanism.

\begin{example}
The continual observation setting is important for public health report publication.
For example, Singapore's Ministry of Health periodically publishes the number of patients
with certain chronic diseases.  Treating each reporting period as a separate and unrelated database, and providing $\epsilon$-differential privacy for each separate database, does not provide $\epsilon$-differential privacy for patients across multiple reporting periods.
To preserve the privacy of patients,
it is important to design mechanisms that can support
periodic release of statistics over a long period of time.
\end{example}

\section{Compressive Mechanism}\label{sec:mechanism}

\subsection{Mechanism Design}

The overall aim of the compressive mechanism is to answer the identity query for databases that have a sparse representation, in an $\epsilon$-differentially private manner.
Algorithm \ref{Algorithm 1} summarizes the compressive mechanism.
The input to the compressive mechanism includes the privacy budget $\epsilon$ and a database $\vD$ 
whose sparse representation is a compressible vector $\vx$ under an orthonormal basis $\vPsi \in \mathbb{R}^{n \times n}$. Vector $\vx$ can be well approximated by its $\vx_S$, for some constant $S$.
$X$ is the set of all databases with a sparse representation.

Applying the sampling operator $\vPhi$ produces a sample vector $\vy = \vPhi\vD \in \mathbb{R}^k$, where $k = \Theta(S\log(n/S))$. We add random noise to each entry of $\vy$. That is to say, for every $\vy[i]$, we have $\vy^*[i] = \vy[i] + \ve[i]$, where $\ve[i] \sim Lap(\sqrt{k}/\epsilon)$. Then $\vy^* = \vy + \ve$.
We can recover a noisy $\vx^*$ from $\vy^*$ through the reconstruction process of compressed sensing. A noisy $\vD^*$ is obtained by $\vD^* = \vPsi \vx^*$. Finally, $\vD^*$ is output by the compressive mechanism.

\begin{algorithm}
\caption{Compressive Mechanism}
{\bf Input}: privacy budget $\epsilon \in \mathbb{R}^+$, $\vD \in X$ (possibly together with a sparse basis $\vPsi \in \mathbb{R}^{n\times n}$).\\
{\bf Output}: $\vD^* \in \mathbb{R}^n$.\\
\label{Algorithm 1}
\begin{algorithmic}[1]
\STATE Generate a (normalized) random matrix $\vPhi \in \mathbb{R}^{k\times n}$ with i.i.d. symmetric Bernoulli distribution.
\STATE Acquire the sample $\vy = \vPhi\vD \in \mathbb{R}^k$.
\STATE Get a noisy sample $\vy^* = \vy + \ve \in \mathbb{R}^k$ for $\ve \in \mathbb{R}^k$ with i.i.d. $Lap(\sqrt{k}/\epsilon)$.
\STATE Reconstruct $\vx^* \in \mathbb{R}^n$.
\STATE Output $\vD^* = \vPsi \vx^*$.
\end{algorithmic}
\end{algorithm}

\subsection{Analysis}

Next we analyze the compressive mechanism. First we prove that it satisfies $\epsilon$-differential privacy.
\begin{lemma}\label{LEM:CORRECT}
The compressive mechanism satisfies $\epsilon$-differential privacy.
\end{lemma}
\proof See the appendix.
\endproof
Next we show that $\vD^*$ is very close to $\vD$, thus ensuring utility.
\begin{lemma}\label{LEM:BOUND}
With high probability, $||\vD - \vD^*||_2 = O(\log(n)/\epsilon)$.
\end{lemma}
\proof See the appendix.
\endproof

The two lemmas above lead immediately to the following theorem.
\begin{theorem}
The compressive mechanism is a universal mechanism with respect to databases with a sparse representation.
\end{theorem}

\subsection{Discussion}
The compressive mechanism is $\epsilon$-differentially private for \emph{all} $\vD \in \mathbb{R}^n$, and works especially well in terms of error bounds for $\vD \in X$. In other words, for any input $\vD \in \mathbb{R}^n$ we guarantee $||\vD^* - \vD||_2 = O(n^{2/3}/\epsilon)$ (from Corollary \ref{generalbound}), while for $\vD \in X$, $||\vD^* - \vD||_2 \le O(\log(n)/\epsilon)$ with high probability. In short, the compressive mechanism is $\epsilon$-differentially private for all $\vD \in \mathbb{R}^n$ and is a universal mechanism for databases with a sparse representation.

Consider the issue of choosing the right $S$, the sparsity parameter. $S$ may not be known in advance, and we may have to choose the best (i.e., with least error) $S$ ourselves. $S$ depends on the input data $\vD$, and since $k = \Theta(S\log(n/S))$ and $||\ve||_2 = O(k/\epsilon)$, $S$ also affects the encoding (adding noise) results. Therefore, we have to choose $S$ in a differentially private way, and a natural method to achieve this aim is the exponential mechanism. $S$ could be any element in $[1, n]$ (if we require the compressive mechanism to work for all possible $\vD \in \mathbb{R}^n$), and for each possible $S$, we define its utility function as
\begin{equation}\label{utility function}
u(\vD, S) = \frac{C_2||\vx - \vx_S||_1}{\sqrt{S}} + \frac{C_4S\log(n/S)}{\epsilon},
\end{equation}
where $C_2$ and $C_4$ are (known) constants. The right hand side of (\ref{utility function}) is the upper bound on the error according to Lemma \ref{CRTresult} (with $\theta$ instantiated). We calculate that the sensitivity of $u(\vD, S), \forall S \in [1, n]$, is $\Delta_{u(\vD, S)} = C_5/\sqrt{S}$, where $C_5$ is a constant. Accordingly, the exponential mechanism outputs $S$ with probability proportional to $e^{-\epsilon u(S)/(2\Delta_{u(S)})}$, where $u(S)$ stands for $u(\vD, S)$; satisfies $\epsilon$-differential privacy; and ensures the near-optimality of $S$ with truly negligible failure probability.

If the total privacy budget is $\epsilon$, we may choose to distribute a small part (say, $0.1\epsilon$ or $0.01\epsilon$) to the exponential mechanism.  In this case, the analysis of the compressive mechanism remains unchanged (recall that we analyze the error in an asymptotic way).

The compressive mechanism may involve the art of identifying a suitable orthonormal basis $\vPsi \in \mathbb{R}^{n\times n}$ under which $\vD \in \mathbb{R}^n$ can be sparse or compressible. This is a mature and profound area in mathematics that has been explored for many years, and we refer the reader to an outstanding book \cite{DL1993}. In this paper, we simply assume that $\vD \in X$ has a sparse representation and its sparse basis $\vPsi \in \mathbb{R}^{n \times n}$ could be treated as part of the input. In fact, many natural data sets have a sparse representation \cite{CW2008}, and that is one reason why compressed sensing has been so widely used since its invention.

The time and space complexity of the compressive mechanism are both $\tilde{O}(n)$.

\section{Continual Observation}\label{sec:continual}
This section focuses on an extension of the compressive mechanism, the \emph{compressive mechanism under continual observation}, or CMCO for short.
\subsection{Mechanism Design}
As in the static case, we suppose that $X \subseteq \mathbb{R}^T$ is the set of all databases with a sparse representation. The overall aim of CMCO is to answer the identity query at each time $t$ in a differentially pan-private manner. The main steps of CMCO are summarized in Algorithm \ref{Algorithm 2}.
	
CMCO takes as input a constant $\epsilon$, the parameter for differential pan-privacy; and (a prefix of) $\vD \in X$, which has a sparse representation $\vx \in \mathbb{R}^T$ under sparse basis $\vPsi \in \mathbb{R}^{T \times T}$. The input differs from the static case in that we do not receive all of $\vD$ at once, but one value at a time; more precisely, at each time $t$, we receive $\vD[t] \in \mathbb{R}$.

At each time $t$, we generate a new random vector $\vPhi_t \in \mathbb{R}^k$ (recall that $k = \Theta(S\log(T/S))$, where $S$ is the sparsity parameter of $\vx$). Each entry of $\vPhi_t$ is distributed according to a symmetric Bernoulli distribution; more concretely,
\begin{align*}
Prob(\vPhi_t[i] = \pm 1/\sqrt{k})= 1/2.
\end{align*}
Upon receiving $\vD[t]$, we apply $\vPhi_t$ to it and get a new vector $\vu_t = \vPhi_t\vD[t] \in \mathbb{R}^k$.

As a result, by time $t$, there are $t$ vectors $\vu_1, \ldots, \vu_t \in \mathbb{R}^k$, forming a matrix $\vM \in \mathbb{R}^{k\times t}$. The $i$-th row of $M$, $i \in [1, k]$, is called $\vm_i \in \mathbb{R}^t$. As $t$ grows, the size of $\vm_i \in \mathbb{R}^t$ grows correspondingly. We apply $k$ independent continual mechanisms to each $\vm_i$ to estimate the sum of the first $t$ entries in $\vm_i \in \mathbb{R}^t$ in an $\epsilon$-differentially pan-private way. The $k$ independent continual mechanisms return a vector $\vv^*_t \in \mathbb{R}^k$ with each entry $\vv^*_t[i]$ representing the estimate of the sum of the first $t$ entries in $\vm_i$. We use $\vv^*_t \in \mathbb{R}^k$ as a resource for reconstruction and obtain a noisy $\vx_t^* \in \mathbb{R}^t$. We obtain $\vD^*_t \in \mathbb{R}^t$ by $\vD^*_t = \vPsi_t\vx^*_t$, where $\vPsi_t \in \mathbb{R}^{t\times t}$ is the orthonormal basis (of the corresponding $\vPsi \in \mathbb{R}^{T\times T}$) in a space of smaller dimension.

We do not store $\vM \in \mathbb{R}^{k \times t}$ at all; and we just discuss it for \emph{analysis}. What we store at each time $t$ is just what is stored for the $k$ $\epsilon$-differentially pan-private continual mechanisms, namely $k$ noisy sums and some independent noise (see the appendix for further details).

At time $t$, CMCO outputs $\vD^*_t \in \mathbb{R}^t$, an estimate of $\vD_t \in \mathbb{R}^t$ (the first $t$ terms of $\vD \in \mathbb{R}^T$).

\begin{algorithm}
\caption{CMCO (Compressive Mechanism under Continual Observation)}
{\bf Initial Input}: privacy budget $\epsilon \in \mathbb{R}^+$. \\
{\bf Input at time $t \in [1,T]$}: $\vD[t] \in \mathbb{R}$.\\
{\bf Output at time $t \in [1,T]$}: $\vD_t^* \in \mathbb{R}^t$.\\
\label{Algorithm 2}
\begin{algorithmic}[1]
\STATE Generate a (normalized) random vector $\vPhi_t \in \mathbb{R}^{k}$ with i.i.d. symmetric Bernoulli distributions.
\STATE Acquire the sample $\vu_t = \vPhi_t\vD[t] \in \mathbb{R}^k$.
\STATE Get an estimate vector $\vv_t^* \in \mathbb{R}^k$, where each $\vv_t^*[i], i \in [1,\ldots,k]$, is the output of a continual mechanism estimating $\sum_{j =1}^t \vu_j[i]$ in an $\epsilon$-differentially pan-private way.
\STATE Reconstruct $\vx^*_t \in \mathbb{R}^t$.
\STATE Output $\vD_t^* = \vPsi_t \vx_t^*$.
\end{algorithmic}
\end{algorithm}

\subsection{Analysis}
In this section, we analyze the performance of CMCO, which needs to be both private and useful.

\begin{theorem}\label{THM:CMCO}
CMCO is $\epsilon$-differentially pan-private and for each time $t \in [1,T]$, $||\vD_t^* - \vD_t||_2 = \tilde{O}(\log^{1.5}(T)/\epsilon)$ with high probability.
\end{theorem}
\proof See the appendix.
\endproof

\subsection{Discussion}
First, what if we directly apply the continual mechanism to a problem? Suppose that at time $t$, we obtain $\Sigma^*_1, \ldots, \Sigma^*_t \in \mathbb{R}$ in an $\epsilon$-differentially pan-private way. Then we perform subtraction operations to get the desired output:
\begin{align*}
\vD_t^*[i] = \Sigma^*_i - \Sigma^*_{i - 1}, \forall i \in [2, t].
\end{align*}
Obviously $\vD_t^*[1] = \Sigma^*_1$. Calculations show that with high probability,
\begin{align*}
||\vD_t - \vD^*_t||_2 = \tilde{O}(\sqrt{t}\log^{1.5}(T)/\epsilon),
\end{align*}
indicating the performance is not as good as that of CMCO. The bounds comparison of CMCO and the continual mechanism is listed in Table \ref{tab:node2}.

\begin{table}
\centering \caption{Error bounds comparison (in terms of $||\vD_t - \vD^*_t||_2$) for answering the identity query at time $t$.}
\begin{tabular}{c c}
\hline \hline
Mechanism&   Bound \\
\hline
 \vspace{1mm} CMCO & $\tilde{O}(\log^{1.5}(T)/\epsilon)$\\
 \vspace{1mm} Continual Mechanism & $\tilde{O}(\sqrt{t}\log^{1.5}(T)/\epsilon)$  \\
\hline
\end{tabular}\label{tab:node2}
\end{table}

Second, the time and space complexity of CMCO are both $\tilde{O}(t + \log(n))$ at each time $t \in [1,T]$.

Last but not least, due to pan-privacy concerns, we are not allowed to store the original input values.

\section{Experiments}\label{sec:exp}

This section illustrates the effectiveness of the compressive
mechanism and CMCO by experimental results on real data sets. We
employ the data sets \emph{social\_network.txt},
\emph{nettrace.txt}, and \emph{search\_logs.txt}, contributed by
Hay, Rastogi, Miklau and Suciu
\cite{DBLP:journals/pvldb/HayRMS10}. The data set
\emph{social\_network.txt} is a graph derived from friendship
relations in an online social network site; \emph{nettrace.txt} is
said to be collected at a university; and \emph{search\_logs.txt}
comes from search query logs collected between $2004$ and $2010$.
\emph{nettrace.txt} consists of $2^{16} = 65536$ entries, while
there are $2^{15} = 32,768$ elements in \emph{search\_logs.txt}.
\emph{social\_network.txt} has $11342$ entries. We also use a real
data set \emph{tcptrace.txt}, which has $7865$ entries.
\emph{tcptrace.txt} was collected by Berkeley, and contain a
30-day trace of the TCP connections between their local network
and the
Internet\footnote{http://ita.ee.lbl.gov/html/contrib/LBL-CONN-7.html}.

\subsection{Haar Basis}

In this part of the section, we discuss results on the data sets
when employing Haar basis in compressive sensing. Haar basis
forms the commonly used and simplest wavelet transformation.

\subsubsection{Choosing a Good $S$}
During the discussion of the compressive mechanism, we mentioned that the exponential mechanism can be used to choose a good sparsity parameter $S$. In this section, we show the feasibility of the exponential mechanism by running it on the three real data sets \emph{social\_network.txt}, \emph{nettrace.txt} and \emph{search\_logs.txt}. We set the total privacy budget to $\epsilon = 1$ and allot $0.1\epsilon = 0.1$ for the exponential mechanism. The remaining $0.9\epsilon = 0.9$ will be used to run the compressive mechanism. In addition, we set $\vPsi \in \mathbb{R}^{n \times n}$ to be a Haar basis \cite{haar}.

\begin{figure*}[t]
\centering
\begin{tabular}[t]{c}
\hspace{-12pt}
   \subfigure[\emph{social\_network.txt}]{
      \psfig{figure=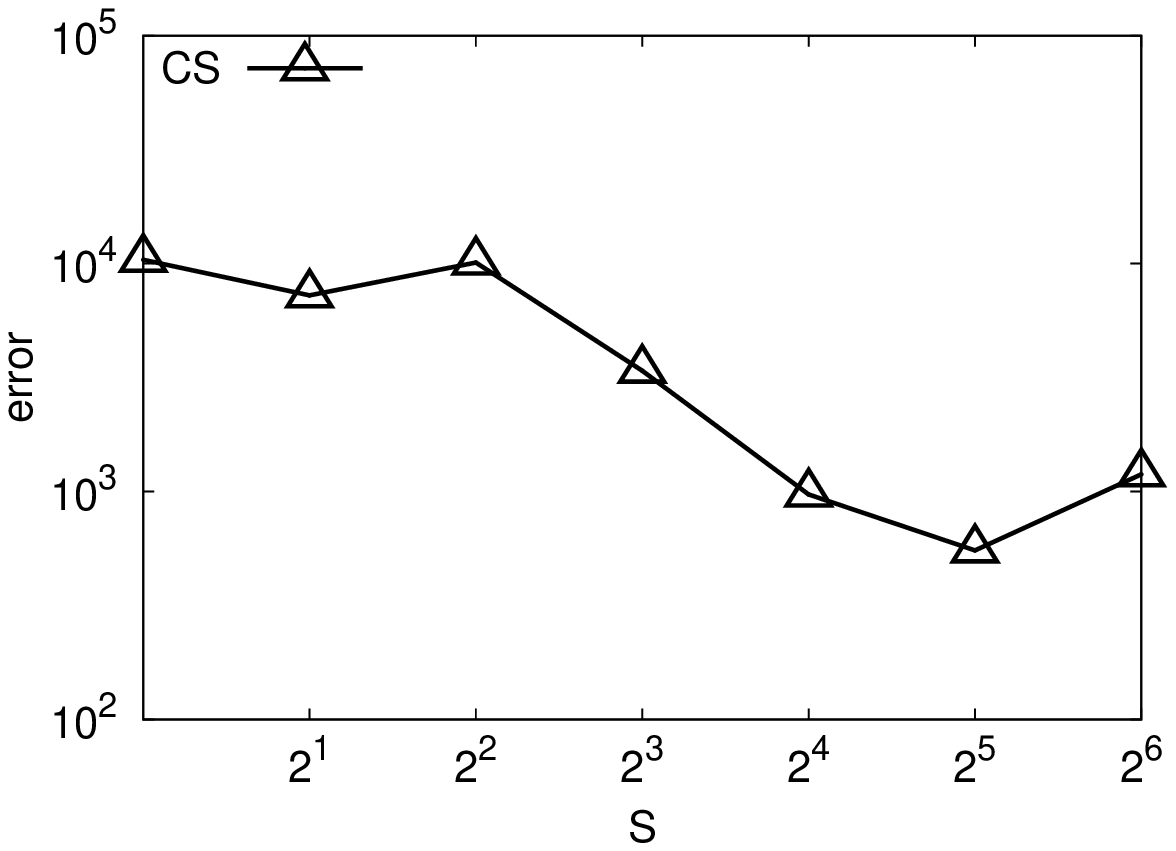,width=.3\linewidth}
      \label{S-error-socialnetwork-1}
   }
\hspace{-12pt}
   \subfigure[\emph{search\_logs.txt}]{
      \psfig{figure=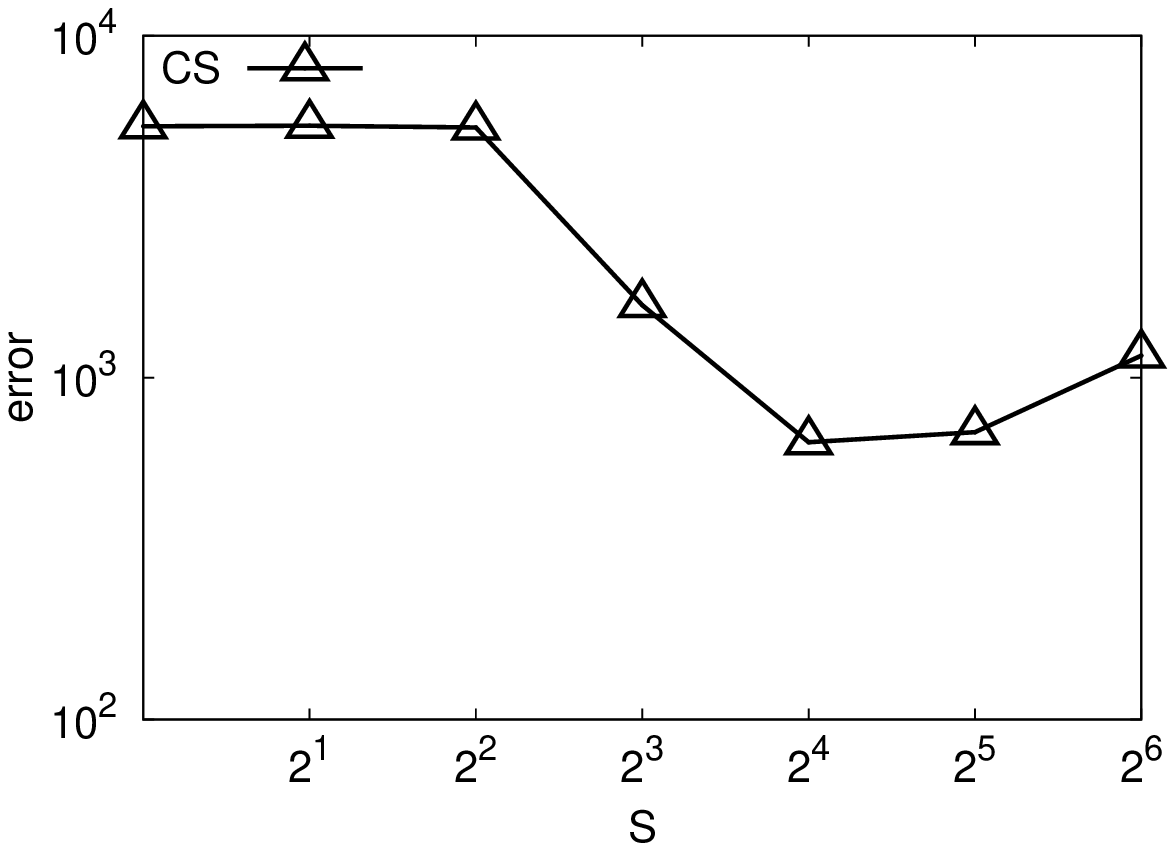,width=.3\linewidth}
      \label{S-error-searchlogs-1}
   }
\hspace{-12pt}
   \subfigure[\emph{nettrace.txt}]{
      \psfig{figure=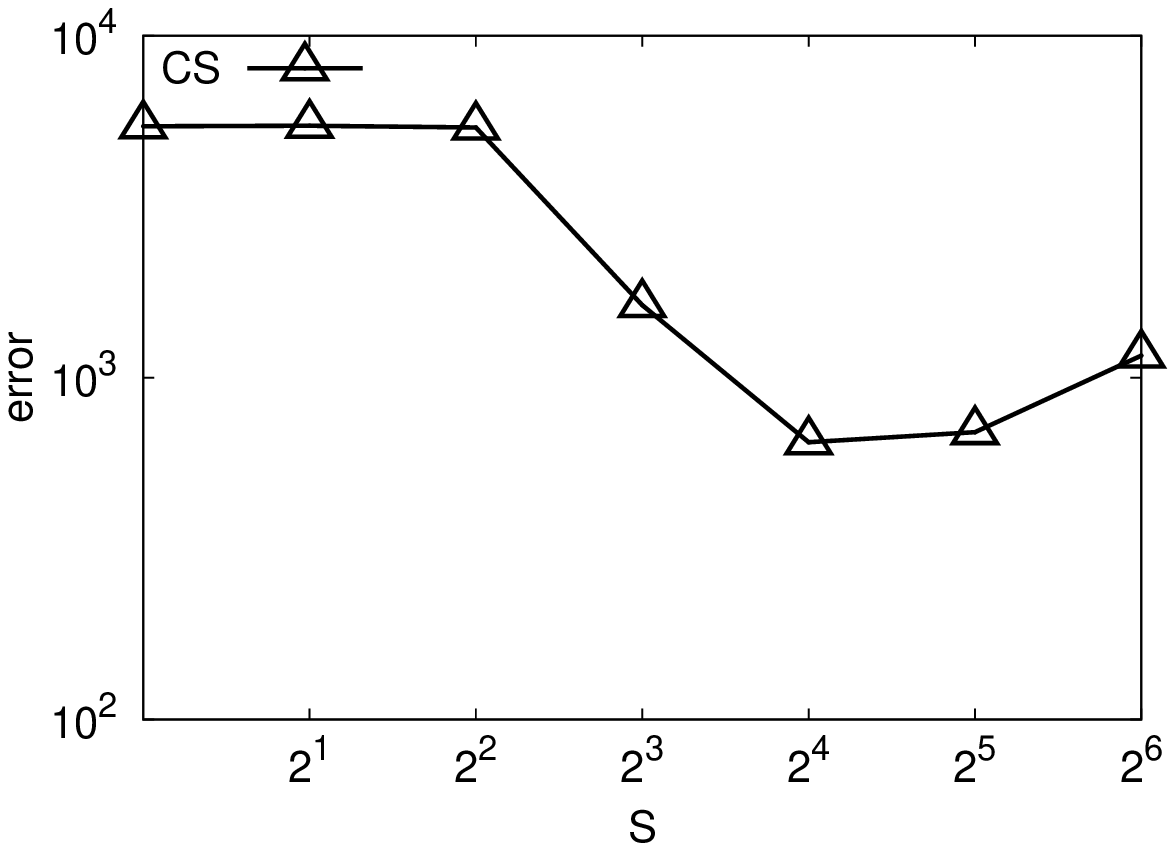,width=.3\linewidth}
      \label{S-error-nettrace-1}
}\\
\hspace{-12pt}
   \subfigure[\emph{social\_network.txt}]{
      \psfig{figure=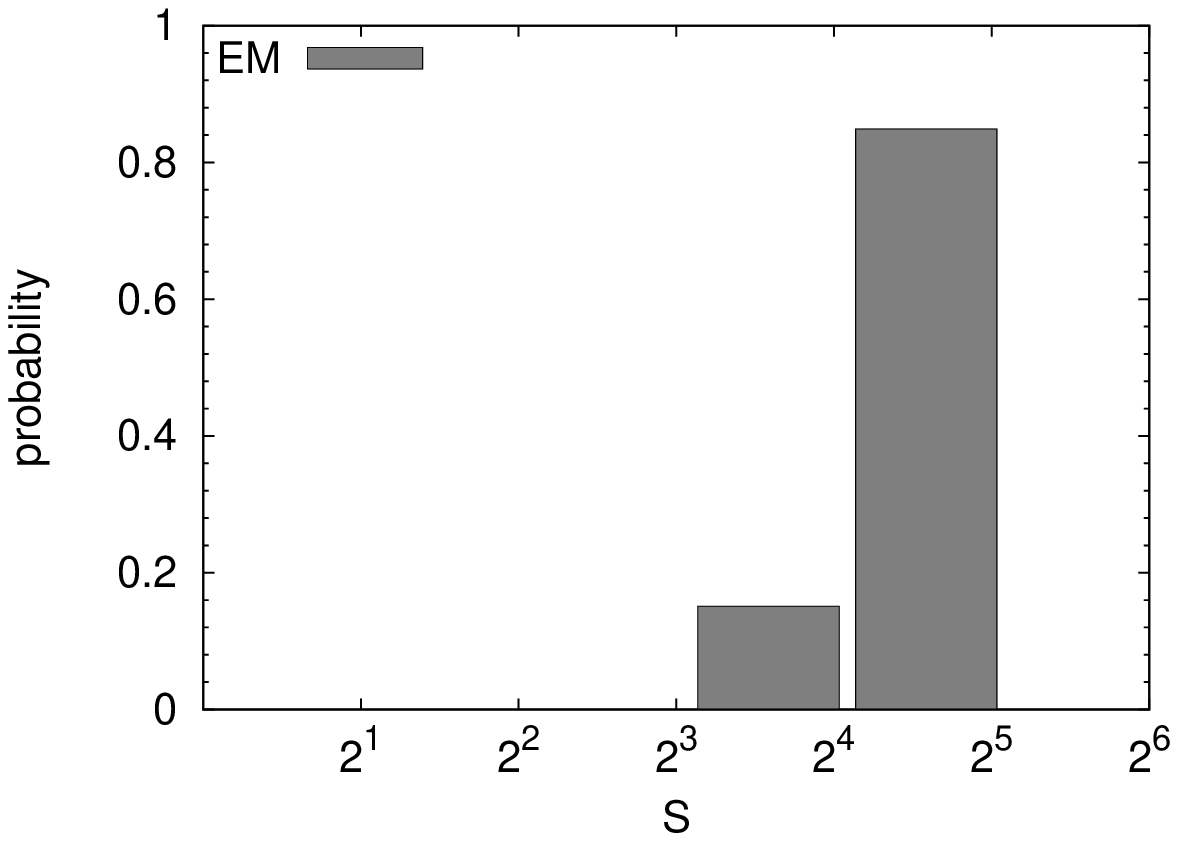,width=.3\linewidth}
      \label{S-error-socialnetwork}
   }
\hspace{-12pt}
   \subfigure[\emph{search\_logs.txt}]{
      \psfig{figure=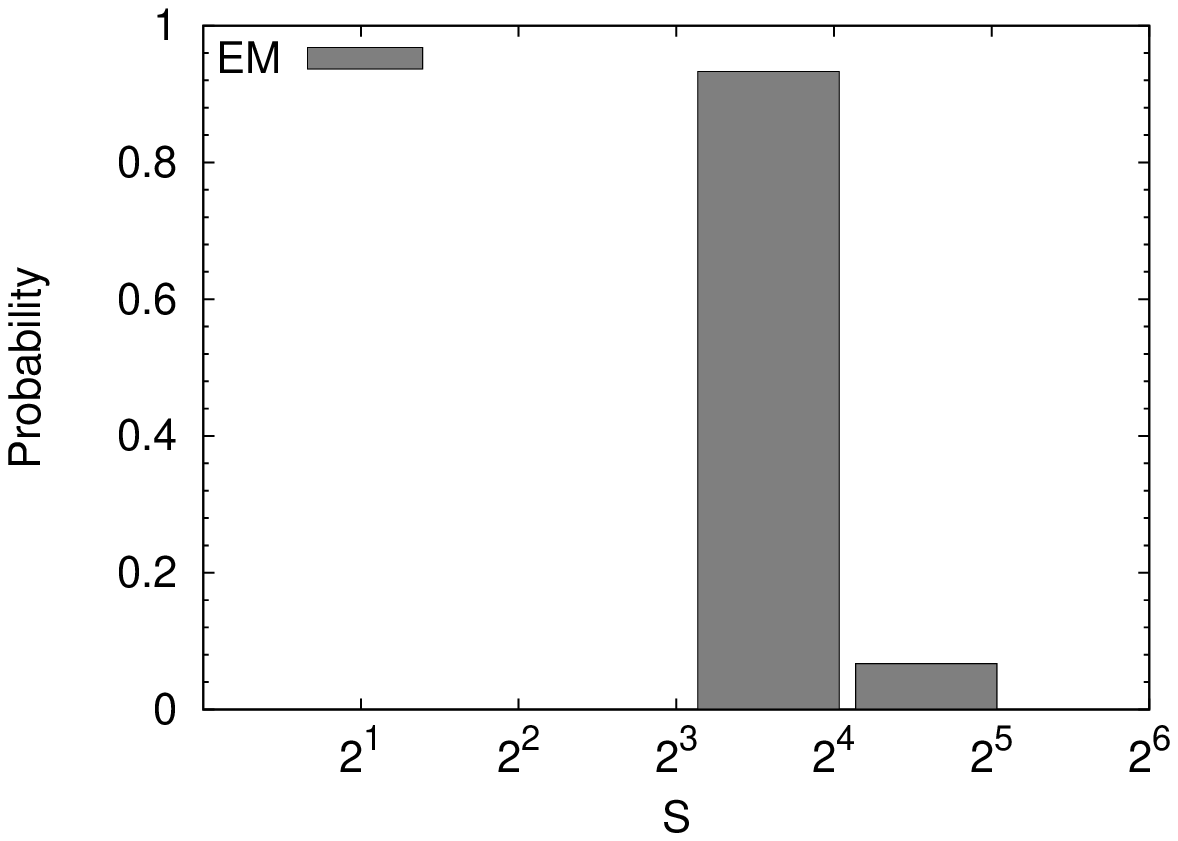,width=.3\linewidth}
      \label{S-error-searchlogs}
   }
\hspace{-12pt}
   \subfigure[\emph{nettrace.txt}]{
      \psfig{figure=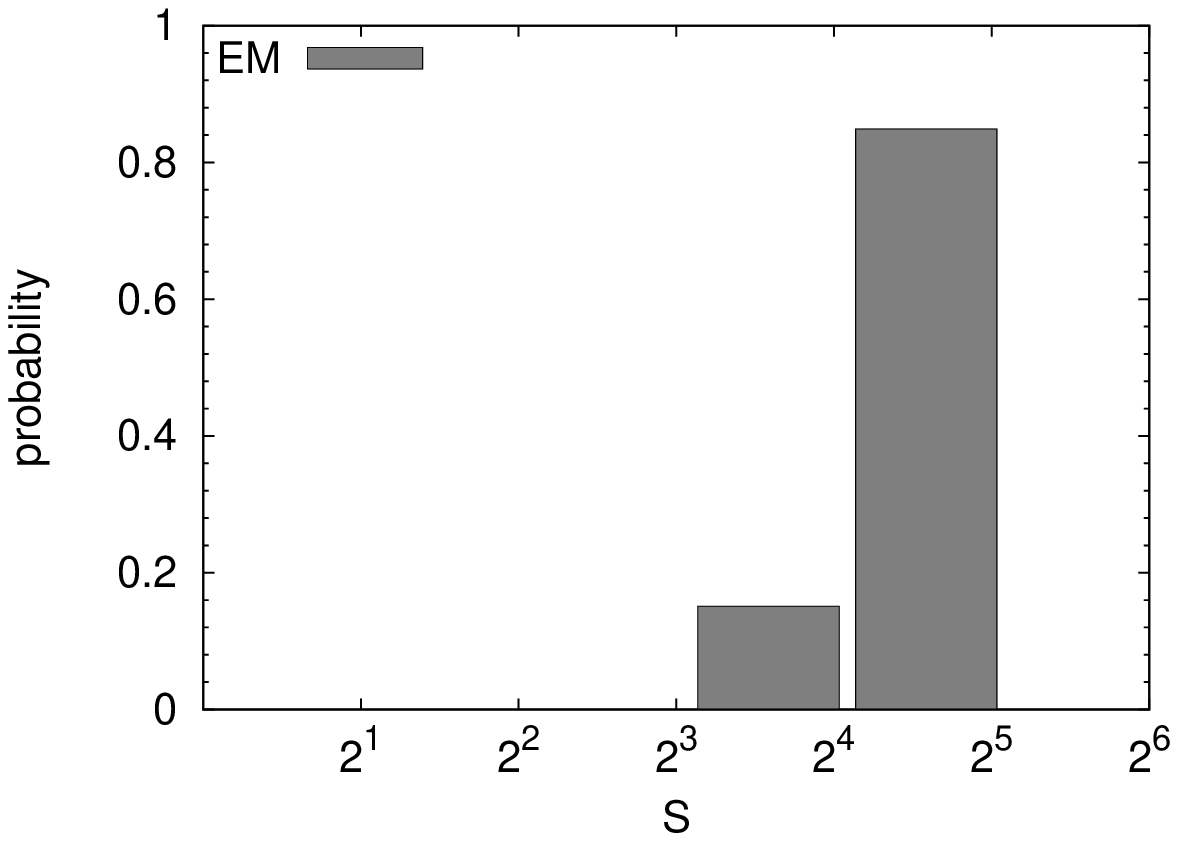,width=.3\linewidth}
      \label{S-error-nettrace}
}
\end{tabular}
\vspace{-5pt}
\caption{\label{Figure 1}Choosing the right $S$: the first three is using CS(Compressive Sensing); the last three is using EM(Exponential Mechanism).}
\vspace{-5pt}
\label{fig: ChooseS}
\end{figure*}

Figure \ref{S-error-socialnetwork-1} through Figure \ref{S-error-nettrace-1} show how error changes as $S$ varies. Then we run the exponential mechanism $1000$ times. Figure \ref{S-error-socialnetwork} through Figure \ref{S-error-nettrace} display the result of the exponential mechanism, explicitly supporting our claim that the exponential mechanism chooses a near-optimal $S$ with truly negligible failure probability.

These experiments show that the exponential mechanism works very well in practice and returns a near-optimal $S$.

\subsubsection{Compressive Mechanism}
In this section, we experimentally evaluate the performance of the compressive mechanism by comparing it with the Laplacian mechanism and HRMS mechanism \cite{DBLP:journals/pvldb/HayRMS10}. The sparse basis is still the Haar basis.

\begin{figure*}[t]
\centering
\begin{tabular}[t]{c}
\hspace{-12pt}
   \subfigure[\emph{social\_network.txt}]{
      \psfig{figure=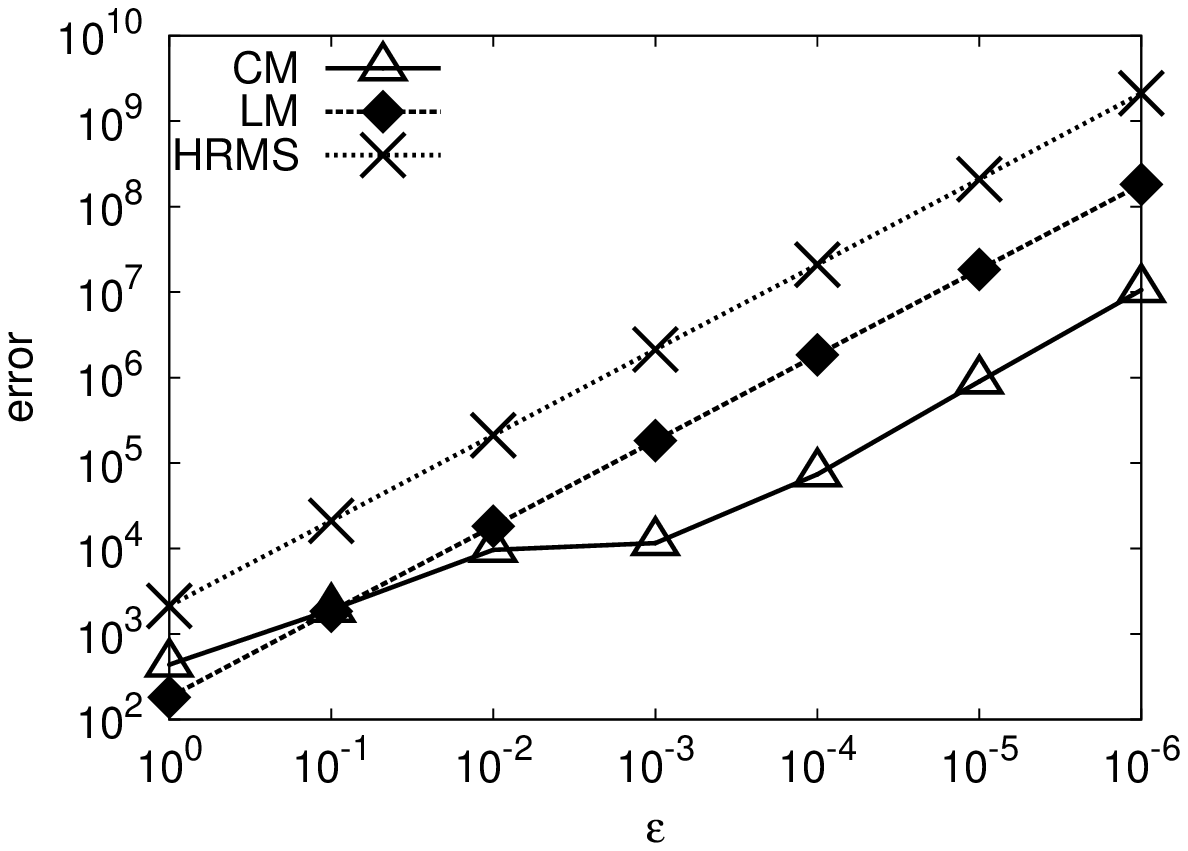,width=.3\linewidth}
      \label{epsilon-error-socialnetwork}
   }
\hspace{-12pt}
   \subfigure[\emph{search\_logs.txt}]{
      \psfig{figure=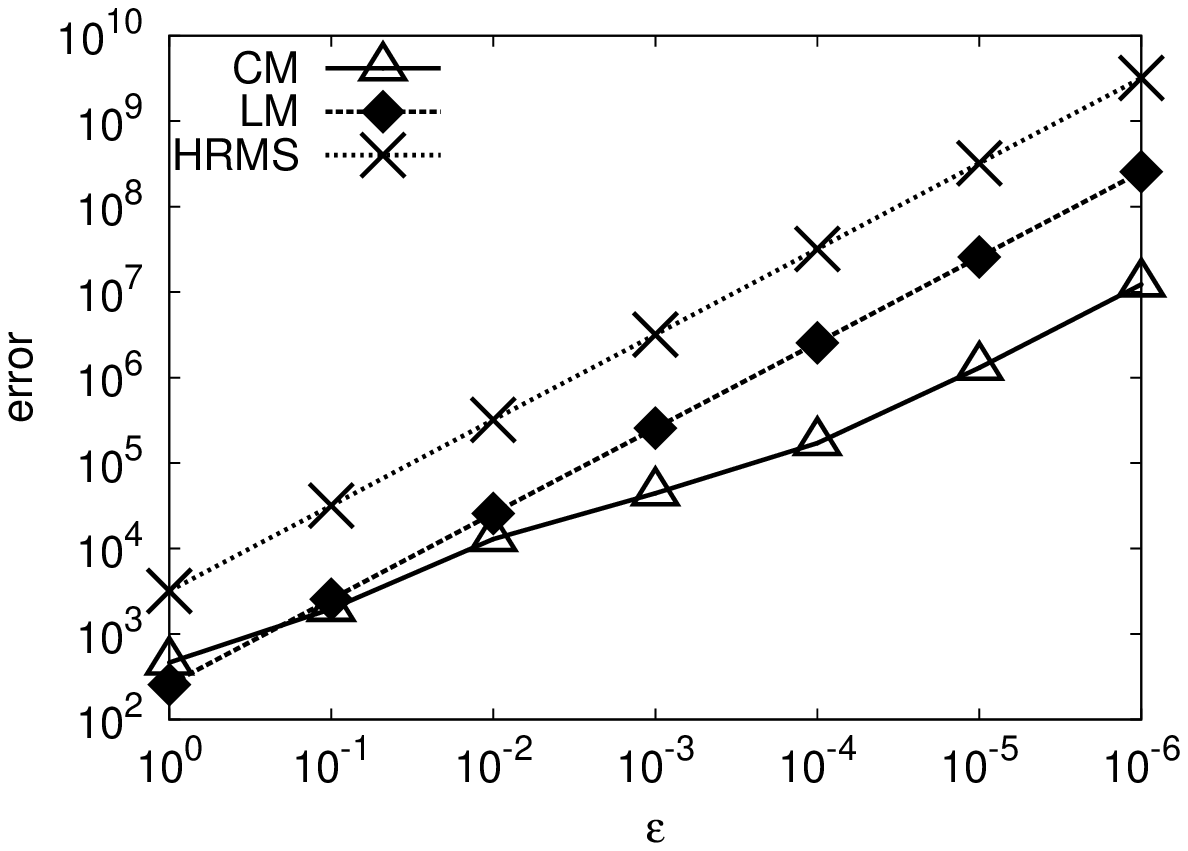,width=.3\linewidth}
      \label{epsilon-error-searchlogs}
   }
\hspace{-12pt}
   \subfigure[\emph{nettrace.txt}]{
      \psfig{figure=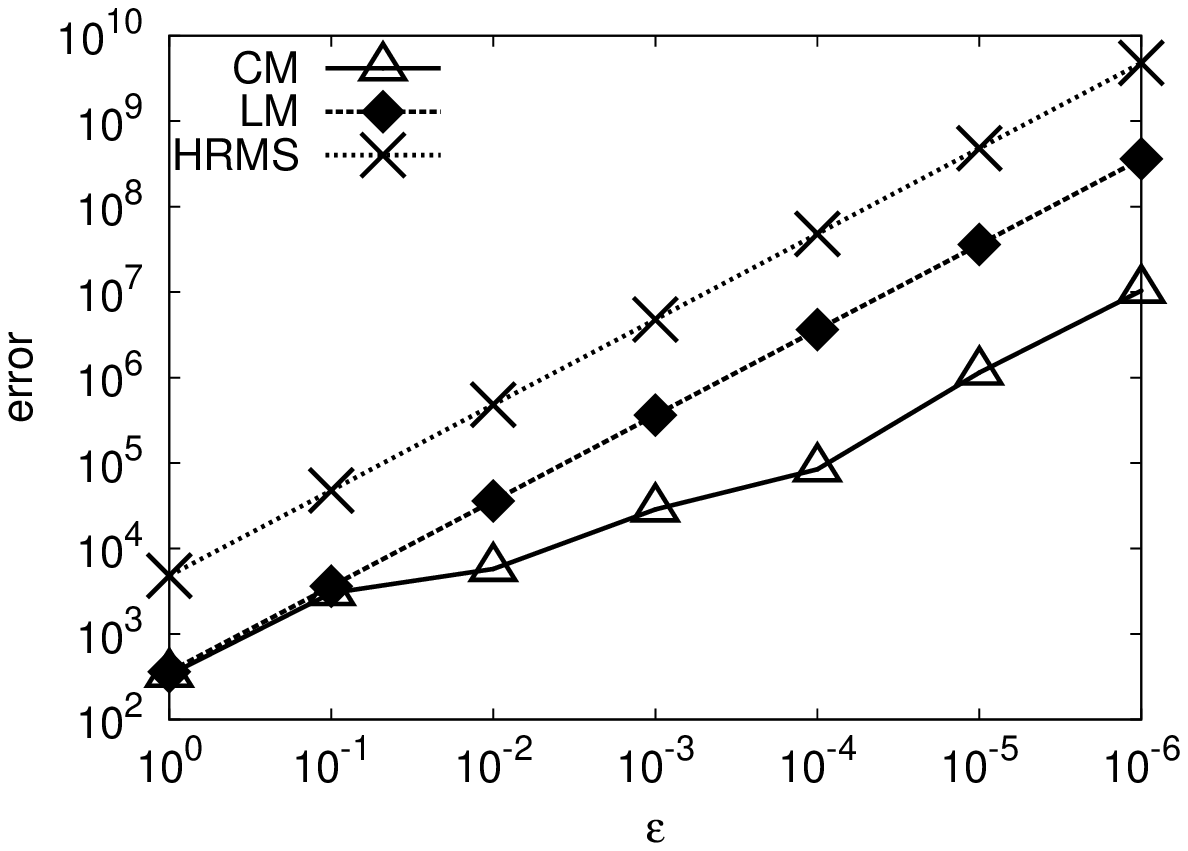,width=.3\linewidth}
      \label{epsilon-error-nettrace}
}
\end{tabular}
\vspace{-5pt}
\caption{\label{Figure 2} A comparison of CM (Compressive Mechanism), LM (Laplacian Mechanism) and HRMS (HRMS Mechanism).}
\vspace{-5pt}
\label{fig: CM}
\end{figure*}

Figure \ref{fig: CM} shows the performance comparison. Both the horizontal and vertical axes use a logarithmic scale. The horizontal axis denotes different choices of the parameter $\epsilon$ for differential privacy; while the vertical coordinates are the errors, namely $||\vD - \vD^*||_2$ of the input $\vD$ and the output $\vD^*$. Overall, the Laplacian and HRMS mechanisms cannot compete with the compressive mechanism, and as $\epsilon$ becomes smaller, the compressive mechanism's advantage becomes larger.

The exception is that for $\epsilon = 1$, the Laplacian mechanism is better than the compressive mechanism. This is because our data set is small, and thus $\sqrt{n}$ is not much larger than $S\log(n/S)$. Combined with the reconstruction errors, the compressive mechanism is slightly worse than the Laplacian mechanism. As $n$ becomes larger, from $11342$ in Figure \ref{epsilon-error-socialnetwork}, $2^{15} = 32768$ in Figure \ref{epsilon-error-searchlogs}, to $2^{16} = 65536$ in Figure \ref{epsilon-error-nettrace}, the advantage of the Laplacian mechanism for the case $\epsilon = 1$ becomes smaller and smaller. For other choices of $\epsilon$, the advantage of the compressive mechanism becomes larger and larger as $n$ increases.

\subsubsection{CMCO}

\begin{figure*}[t]
\centering
\begin{tabular}[t]{c}
\hspace{-12pt}
   \subfigure[$\epsilon = 10^{-1}$, \emph{social\_network.txt}]{
      \psfig{figure=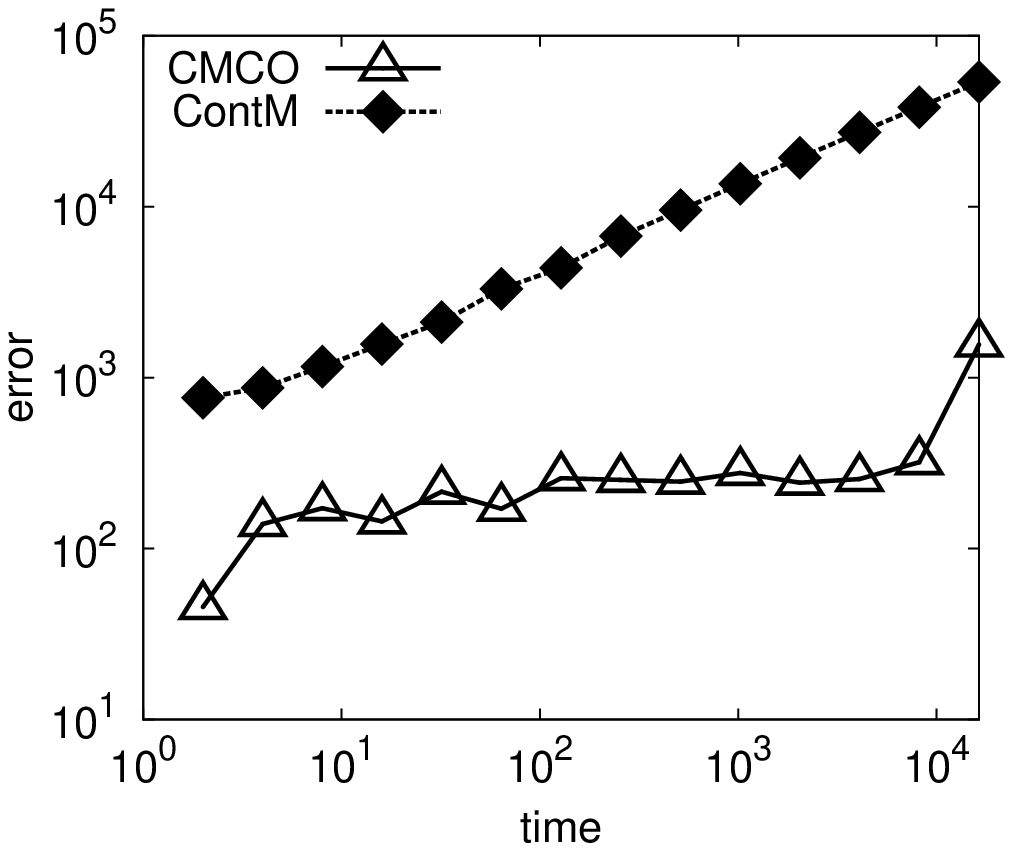,width=.3\linewidth}
      \label{time-error-socialnetwork-0.1}
   }
\hspace{-12pt}
   \subfigure[$\epsilon = 10^{-1}$, \emph{search\_logs.txt}]{
      \psfig{figure=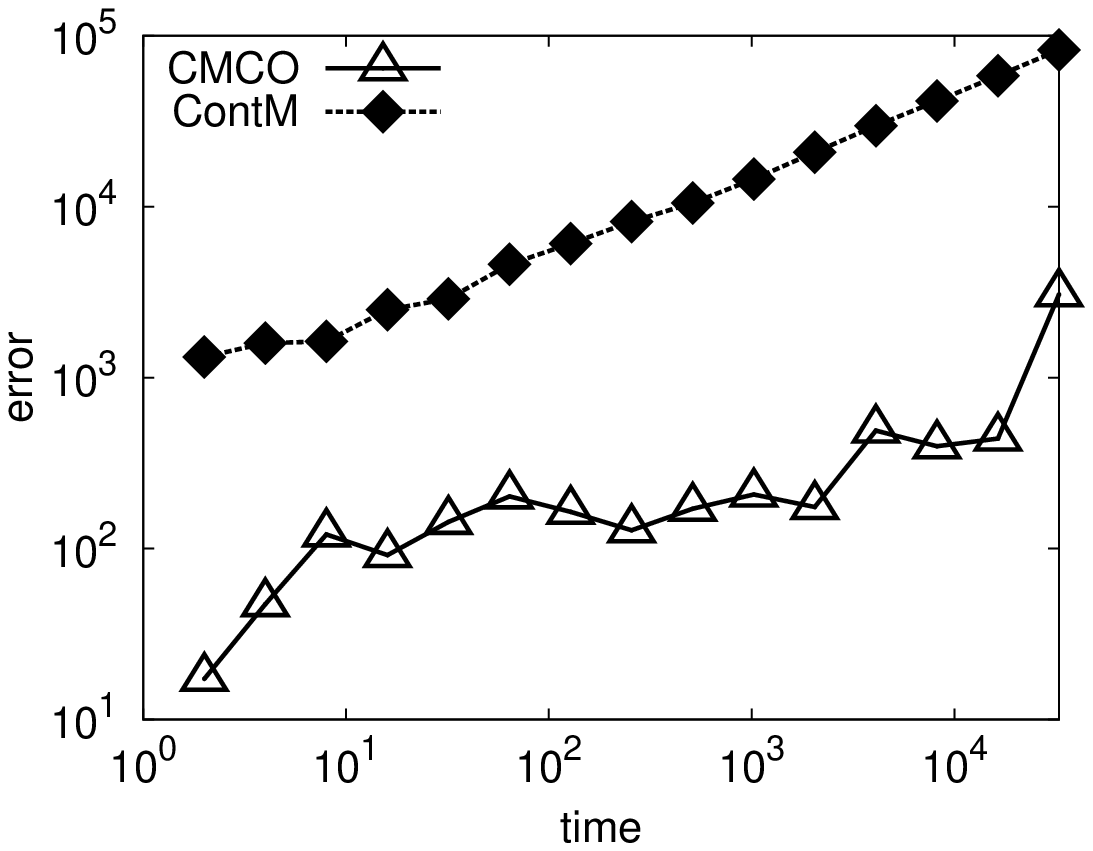,width=.3\linewidth}
      \label{time-error-searchlogs-0.1}
   }
\hspace{-12pt}
   \subfigure[$\epsilon = 10^{-1}$, \emph{nettrace.txt}]{
      \psfig{figure=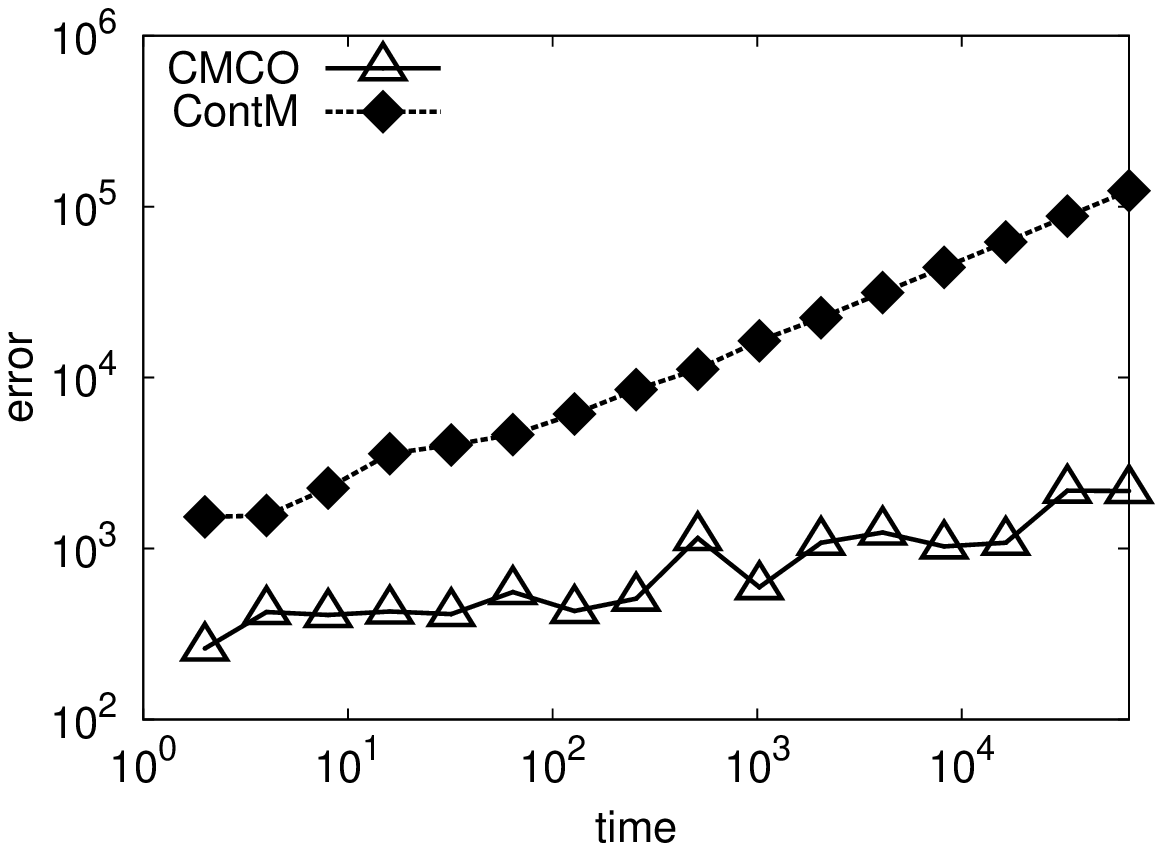,width=.3\linewidth}
      \label{time-error-nettrace-0.1}
}\\
\hspace{-12pt}
   \subfigure[$\epsilon = 10^{-3}$, \emph{social\_network.txt}]{
      \psfig{figure=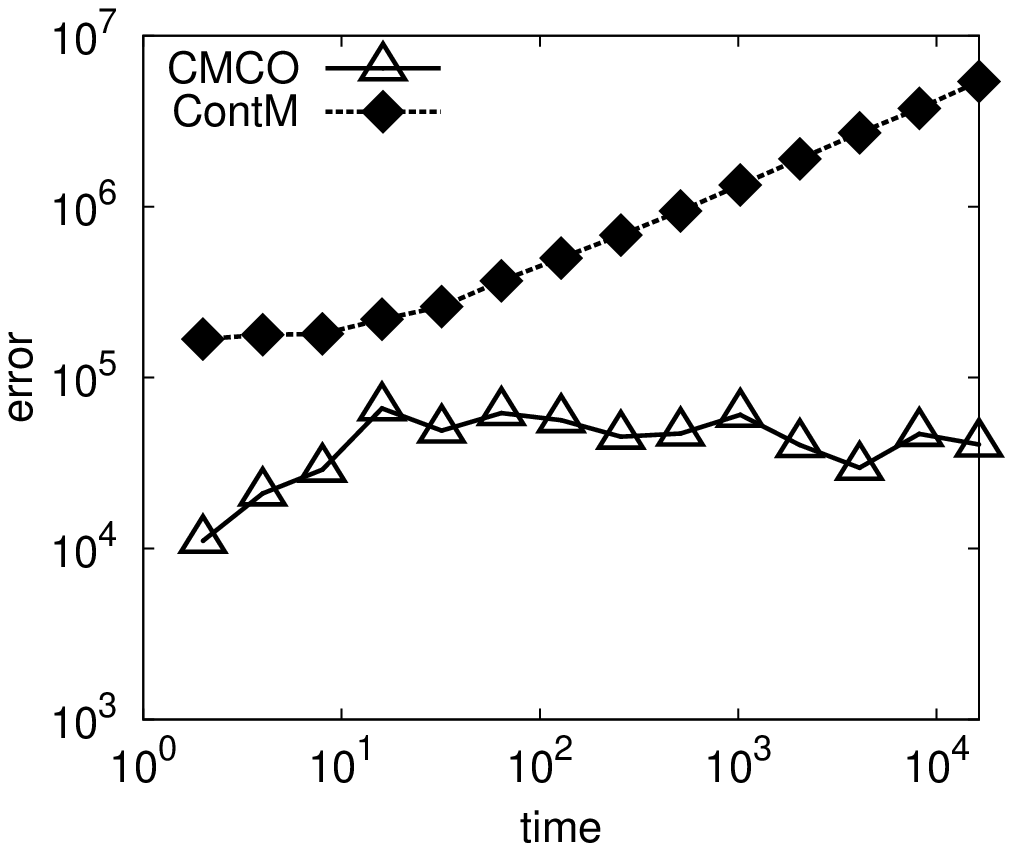,width=.3\linewidth}
      \label{time-error-socialnetwork-0.001}
}
\hspace{-12pt}
   \subfigure[$\epsilon = 10^{-3}$, \emph{search\_logs.txt}]{
      \psfig{figure=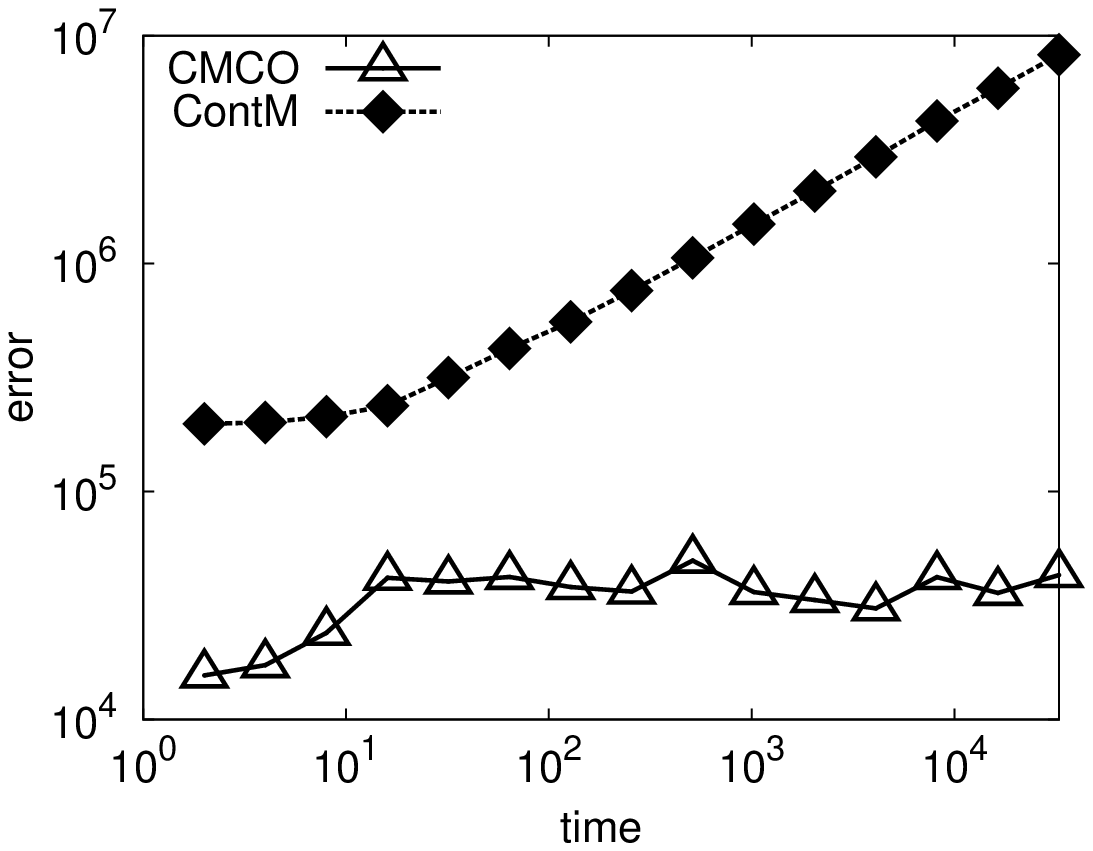,width=.3\linewidth}
      \label{time-error-searchlogs-0.001}
}
\hspace{-12pt}
   \subfigure[$\epsilon = 10^{-3}$, \emph{nettrace.txt}]{
      \psfig{figure=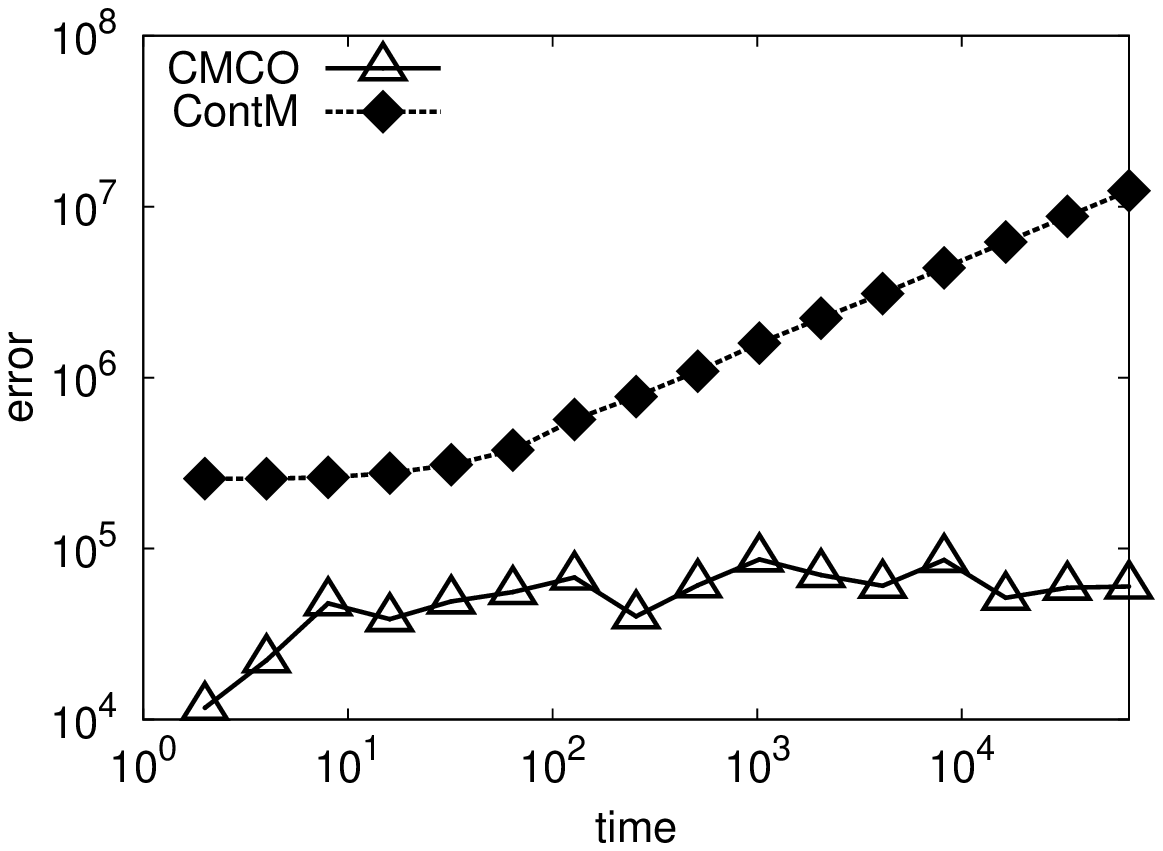,width=.3\linewidth}
      \label{time-error-nettrace-0.001}
}\\
\hspace{-12pt}
   \subfigure[$\epsilon = 10^{-5}$, \emph{social\_network.txt}]{
      \psfig{figure=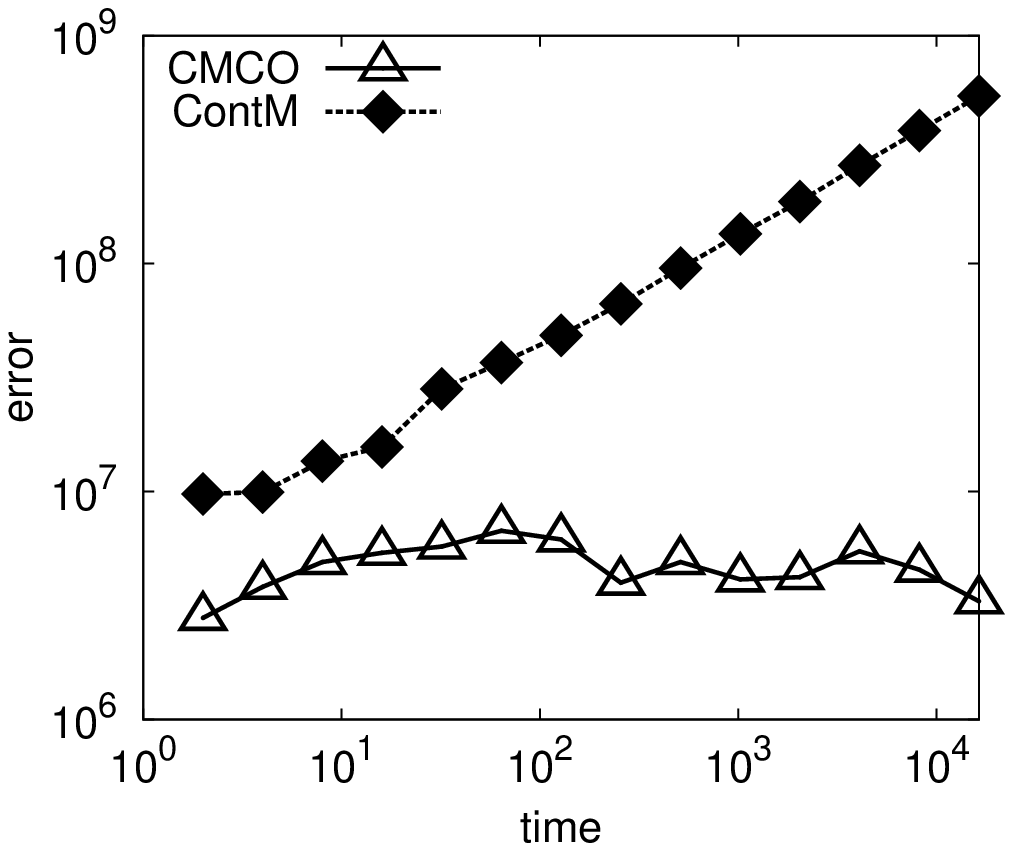,width=.3\linewidth}
      \label{time-error-socialnetwork-0.00001}
}
\hspace{-12pt}
   \subfigure[$\epsilon = 10^{-5}$, \emph{search\_logs.txt}]{
      \psfig{figure=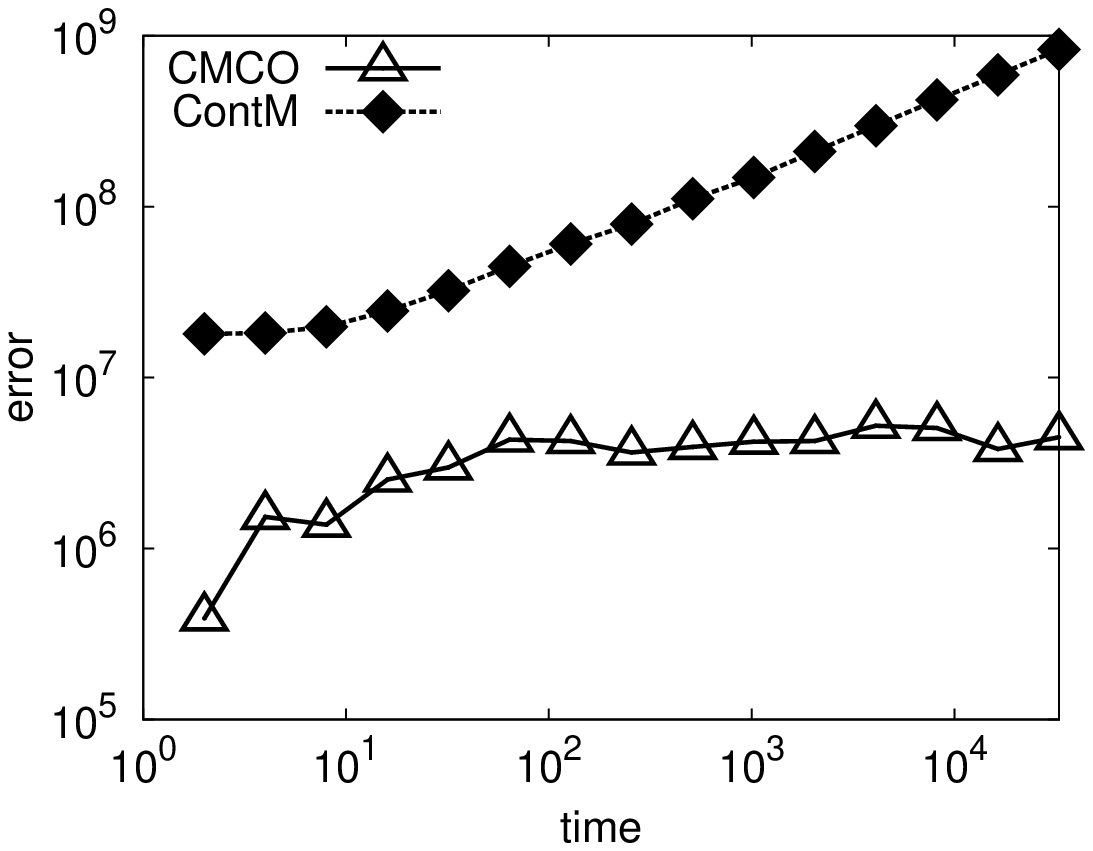,width=.3\linewidth}
      \label{time-error-searchlogs-0.00001}
}
\hspace{-12pt}
   \subfigure[$\epsilon = 10^{-5}$, \emph{nettrace.txt}]{
      \psfig{figure=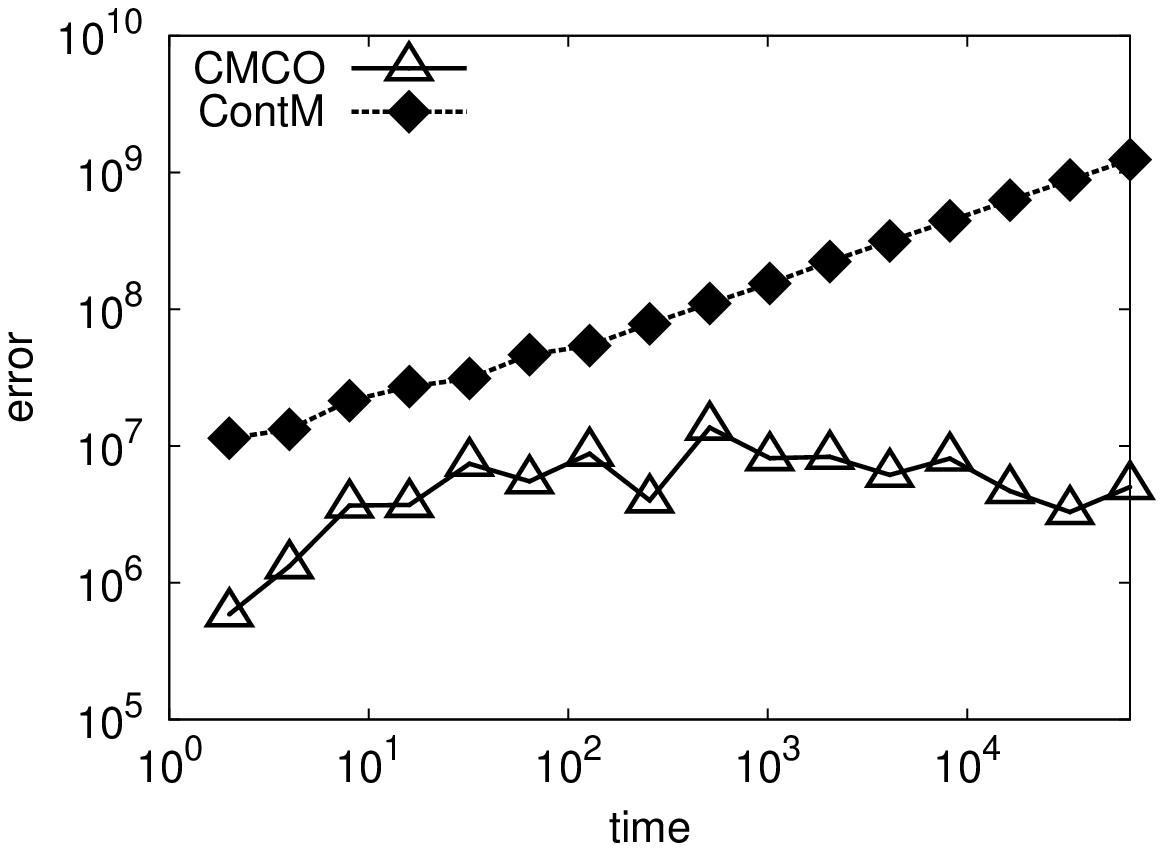,width=.3\linewidth}
      \label{time-error-nettrace-0.00001}
}
\end{tabular}
\vspace{-5pt}
\caption{\label{Figure 3} A comparison of CMCO (Compressive Mechanism under Continual Observation) and ContM (Continual Mechanism).}
\vspace{-5pt}
\label{fig: CMCO}
\end{figure*}

To demonstrate the strength of CMCO over the continual mechanism, we set $\epsilon$ and run the two algorithms on \emph{social\_network.txt}, \emph{search\_logs.txt} and \emph{nettrace.txt}. Figure \ref{fig: CMCO} shows that the continual mechanism has significantly lower utility than the compressive mechanism under continual observation. The $x$-axis varies the time $t$ and the $y$-axis is the corresponding change in error, namely $||\vD_t - \vD_t^*||_2$. The sparse basis is still the Haar basis.

\subsection{Cosine Basis}

\begin{figure}[t]
\centering
\begin{tabular}[t]{c}
\hspace{-12pt}
   \subfigure[\emph{tcptrace.txt}]{
      \psfig{figure=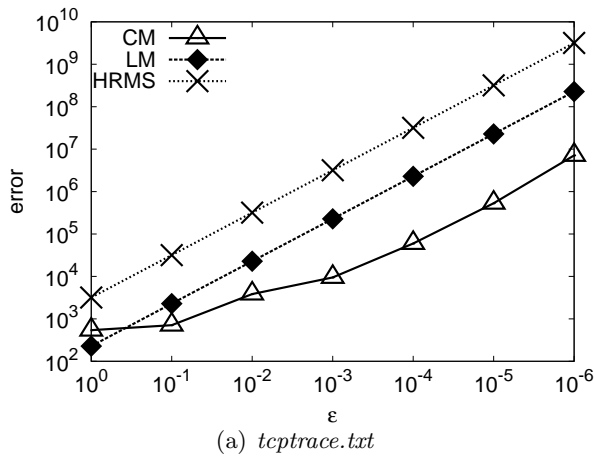,width=0.5\linewidth}
      \label{epsilon-error-tcptrace}
   }
\end{tabular}
\vspace{-5pt}
\caption{\label{Figure 4} A comparison of CM (Compressive Mechanism), LM (Laplacian Mechanism) and HRMS (HRMS Mechanism) for the Cosine Basis.}
\vspace{-5pt}
\label{fig: CosineBasis}
\end{figure}

The previous experiments used the Haar basis as the orthonormal basis $\vPsi \in \mathbb{R}^{n \times n}$. This is an appropriate choice because the three data sets of Hay, Rastogi, Miklau and Suciu \cite{DBLP:journals/pvldb/HayRMS10} have the property that every two adjacent elements in a file are very close to each other. To illustrate the power of the compressive mechanism, here we employ a different orthonormal basis, the cosine basis \cite{cosine}, on a time-related data set \emph{tcptrace.txt}. Figure \ref{fig: CosineBasis} shows that the compressive mechanism far outperforms the Laplacian mechanism and HRMS mechanism.

\section{Concluding Remarks}\label{sec:conclu}
We have introduced  the compressive mechanism as a means of realizing the idea of a universal mechanism. We have provided theoretical bounds and experimental results for the compressive mechanism, and showed how to apply the compressive mechanism to a case of continual observation.

As mentioned earlier, one open problem concerns the lower bound of error in the definition of the universal mechanism. Formally, assume that there is an $\epsilon$-differentially private mechanism to answer the identity query. Then what is the lower bound of $||\vD - \vD^*||_2$, where $\vD^* \in \mathbb{R}^n$ is the output of the mechanism? Here $X$ can be a subset of $\mathbb{R}^n$, such as databases with a sparse representation.

Regarding future work: are there applications of the compressive mechanism to other scenarios beyond continual observation? What if we use the compressive mechanism to satisfy other privacy definitions such $(\epsilon, \delta)$-differential privacy or zero-knowledge privacy? How can one incorporate considerations such as integrity and consistency constraints into our framework?

{\small \bibliography{csdp,privacy}}
\bibliographystyle{abbrv}

\clearpage

\appendix
\section{Proof of Lemma \ref{LEM:CORRECT}}
First let us focus on the sampling process, which can be characterized as a random projection $g: X \rightarrow \mathbb{R}^k$. For any $\vD \in X$, $g(\vD) = \vy = \vPhi\vD$. We have mentioned that the random matrix $\vPhi$ is formed
by sampling i.i.d. entries from a symmetric Bernoulli distribution; more accurately,
\begin{align*}
Prob(\vPhi(i,j) = \pm 1/\sqrt{k}) = 1/2.
\end{align*}
The sensitivity of $g$ is $\Delta_g = \sqrt{k}$. In the compressive mechanism, we make use of a Laplacian mechanism $\mathcal{K}_g: X \rightarrow \mathbb{R}^k$, which is $\epsilon$-differentially private according to Lemma \ref{LaplaceMechanism}.

The subsequent reconstruction process of compressed sensing is a deterministic process and does not involve probability. Thus the whole compressive mechanism is $\epsilon$-differentially private.

\section{Proof of Lemma \ref{LEM:BOUND}}
Each $\ve[i]$ is distributed according to
\begin{align*}
Lap(\sqrt{k}/\epsilon) = Lap(\lambda).
\end{align*}
We define a random variable $Y = \sum_{i = 1}^k \ve[i]^2$. We can compute that $E(Y) = 2k\lambda^2$ and that $Var(Y) = 20k\lambda^4$. By the Chebyshev inequality\footnote{The Chernoff bound cannot be applied, as the integral defining the moment-generating function of $Y$ does not converge unconditionally in a neighborhood of $0$.},
\begin{align*}
Prob(|Y| \le 2\lambda^2(k + \sqrt{2k/\delta})) \ge 1 - \delta.
\end{align*}
Along with the fact that $||\ve||_2 = \sqrt{Y}$, we see that with probability at least $1 - \delta$,
\begin{align*}
||\ve||_2 = O(k/(\epsilon \delta^{1/4})).
\end{align*}
Recall that with probability at least $1 - \exp(-k)$, $\vA = \vPhi\vPsi \in \mathbb{R}^{k \times n}$ satisfies RIP. Using a union bound and Corollary \ref{ErrorBound}, with high probability,
\begin{align*}
||\vD - \vD^*||_2 = O(\log(n)/\epsilon).
\end{align*}

\section{Proof of Theorem \ref{THM:CMCO}}

The $\epsilon$-differential pan-privacy of CMCO is straightforward: at time $t$, we estimate $\sum_{j =1}^t \vu_j[i]$ in an $\epsilon$-differentially pan-private way and any operation after this step preserves pan-privacy.

Suppose that at each time $t$, each entry $\vv_t[i]$ of the vector $\vv_t \in \mathbb{R}^k$ represents the sum of the first $t$ entries in $\vm_i \in \mathbb{R}^t$. We apply a continual mechanism to each $\vm_i \in \mathbb{R}^t$. For two neighboring inputs $\vD_t, \vD'_t \in \mathbb{R}^t$ with $||\vD_t - \vD'_t||_1 \le 1$, their corresponding $\vm_i, \vm'_i \in \mathbb{R}^t$ satisfies $||\vm_i - \vm'_i||_1 \le 1/\sqrt{k}$. Combining this fact with Lemma \ref{continual}, we know that with probability at least $1 - \beta$,
\begin{align*}
|\vv_t[i] - \vv^*_t[i]| = O(\log(1/\beta)\log^{1.5}(T)/(\epsilon\sqrt{k})),
\end{align*}
for each $i \in [1,\ldots,k]$. By a union bound, with probability at least $1 - \beta$,
\begin{align*}
|\vv_t[i] - \vv^*_t[i]| = O(\log(k/\beta)\log^{1.5}(T)/(\epsilon\sqrt{k}))
\end{align*}
holds for all $i$ simultaneously. Along with the Chebyshev inequality, we have that with probability at least $1 - \beta - \delta$,
\begin{align*}
||\vv_t - \vv^*_t||_2 = O(\log(k/\beta)\log^{1.5}(T)/(\epsilon\delta^{1/4})).
\end{align*}
Then by Corollary \ref{ErrorBound} and a union bound, we obtain the result that with high probability,
\begin{align*}
||\vD_t - \vD^*_t||_2 = \tilde{O}(\log^{1.5}(T)/\epsilon).
\end{align*}

\section{More on Reconstruction Process}
In the paper, we just used the reconstruction process in a black-box manner. Here we discuss the two reconstruction algorithms mentioned \cite{CRT2006, DBLP:journals/cacm/NeedellT10} more concretely.

Remember the equation we have mentioned:
\begin{equation}\label{appen}
\vy^* = \vA\vx + \ve,
\end{equation}
where $\vA \in \mathbb{R}^{k \times n}$, $\ve, \vy^* \in \mathbb{R}^k$ and $\vx \in \mathbb{R}^n$. $\vA$ and $\vy^*$ are known, $\ve$ is some unknown noise and we want to know what $\vx$ is (reconstruct $\vx$).

What \cite{CRT2006} does is to solve the convex program
\begin{align*}
\min_{\vx^* \in \mathbb{R}^n} ||\vx^*||_1 \qquad s.t. ||\vA\vx^* - \vy||_2 \le \theta,
\end{align*}
where $\theta$ bounds the amount of noise $\ve$.

\cite{DBLP:journals/cacm/NeedellT10} is more complicated by using a greedy algorithm to solve (\ref{appen}). The essential idea is to iteratively approximate the target vector $\vx$. At each step, the current approximation induces a residual, the part of the target vector that has not been approximated. Samples are updated so that they reflect the current residual and are used to construct a proxy for the residual, allowing us to identify the large components in the residual. This step yields a tentative support for the next approximation and samples are used to estimate the approximation on this support set using least squares. The process is repeated until recoverable energy is met in the vector.
\section{More on Continual Mechanism}
We also used continual mechanism \cite{DBLP:conf/stoc/DworkNPR10} in a black-box way in the body of the paper. Here we discuss the content of the mechanism.

Continual mechanism receives an input $\vsigma(t) \in \{0, 1\}$ at each time $t \in [T]$, and outputs an approximation to the number of $1$'s seen in the length $t$.

There is a preprocessing step: segmentation. For $i \in [\log T]$ (assume $T$ to be a power of $2$ without loss of generality), associate with each string $s \in \{0, 1\}^i$ the time segment $S$ of $2^{\log T - i}$ time periods $\{s\circ 0^{\log T - i}, \ldots, s \circ 1^{\log T - i}\}$. The segment begins in time $s \circ 0^{\log T - i}$ and ends in time $s \circ 1^{\log T - i}$.

Then comes the processing steps. At each time $t \in [T]$, continual mechanism maintains $\log T$ segments that contain $t - 1$. Each of these $\log T$ segments has a noise associated with it sampling from $Lap((1+ \log T)/\epsilon)$. The output at each time $t \in [T]$ is the summation of the noise from these $\log T$ segments and the count (true count plus a random noise sampling from $Lap((1+ \log T)/\epsilon)$).

\end{document}